\documentclass[twocolumn,aps,prd,tighten,amssymb,,amsmath,showpacs]{revtex4}
\usepackage{amssymb}
\usepackage{epsfig}

\begin{document}

\title{Head-on collisions of boson stars}

\author{ C.~Palenzuela$^{1}$, I.~Olabarrieta$^{1,2}$, L.~Lehner$^{1}$, S.~Liebling$^{3}$}

\affiliation{$1$ Department of Physics and
Astronomy, Louisiana State University, 202 Nicholson Hall, Baton
Rouge, Louisiana 70803-4001, USA\\
$2$ TELECOM Unit, ROBOTIKER-Tecnalia, Ed. 202 Parque Tecnol\'ogico Zamudio E-48170, Bizkaia, SPAIN\\
$3$ Department of Physics, Long Island University -- C.W. Post Campus,
Brookville, New York 11548, USA}

\begin{abstract}
We study head-on collisions of boson stars in three dimensions.
We consider evolutions of two boson stars which may differ in their phase or have
opposite frequencies but
are otherwise identical. Our studies show that these phase differences result
in different late time behavior and gravitational wave output.
\end{abstract}

\maketitle

\section{Introduction}
\label{introduction}
Models of compact ``stars'' described by non-fluid sources serve both
as probes of strong field gravity
as well as exotic alternatives to the standard ``regular'' fluid stars.
Among these non-standard star models, a geon --a self
gravitating star consisting of electromagnetic fields-- was first considered 
by Wheeler~\cite{wheeler}. 
Despite its appeal, no stable geon could be constructed and so interest diminished.
However, borrowing from this idea, Kaup was able to construct the
first stable configuration --referred to as a {\it Klein-Gordon geon}-- 
by considering a $U(1)$ classical scalar field~\cite{kaup}.
Soon after Kaup's construction, Ruffini and Bonazzola revisited
the solution, adopting a quantum point of view for the matter fields
though maintaining the geometry as a classical entity~\cite{ruffini}. By
employing a Hartree-Fock (multi-particle) approximation, they re-obtain Kaup's equations
(and hence solutions) providing an alternative view to the resulting stars.
These solutions are currently known as {\em boson stars} and describe a family of
self-gravitating scalar field configurations within general relativity.

Boson stars have been studied in many different contexts (see
\cite{jetzer},\cite{mielke} for reviews). They have been proposed
as models for dark matter (see \cite{dark_matter} for a recent review) 
and more recently have been used as convenient models for
compact objects including black holes \cite{guzman06}. 
One can consider strongly gravitating compact-star spacetimes 
without the worries associated with regular matter --such as shock
fronts and discontinuities in the fluid variables-- making boson stars very useful 
probes of strong-field general relativity.

Given this utility, a number of efforts have modeled the dynamical evolution of boson stars.
For instance, they have been considered in
one and two dimensions for studies of critical phenomena~\cite{hawley,lai}.
In three dimensions, studies of
the dynamics of perturbed boson stars have been presented in \cite{guz04,BBDGS05}
and preliminary collisions in \cite{balakrishna_phd}. 
Works examining boson star collisions in two dimensions were carried out
in \cite{lai} while in three dimensions, in addition to our own, another effort is
under way~\cite{frans}.
Another work which studies the collisions of Q-balls also has relevance here~\cite{Qball}.
These Q-balls are solitonic solutions of the Klein-Gordon equations in the
absence of gravity with an attractive potential.

In this paper we investigate several boson star configurations paying particular attention to
head-on collisions in 3D. Our aim is to survey some of the phenomenology that can be found 
in these scenarios. Additionally, since one can employ this particular 
physical system to describe strongly gravitating compact stars, one can use them to model
compact systems in general relativity in their early stages where tidal effects
are not too severe.
On a more speculative level, one can also examine 
the gravitational wave signature produced  so that searches
by gravitational wave detectors might place constraints on their existence.

The paper is organized as follows. In Section \ref{equations} we detail
the formalism we use for the Einstein-Klein-Gordon system.
Section \ref{numerics} describes the numerical implementation of the governing equations. 
In Section IV we present our results.
We conclude in Section~\ref{conclusions} adding some final comments.
The computation of the initial data for a single boson star is described in
the Appendix~\ref{initial_data}, while the Appendix~\ref{energy_considerations} is devoted
to some (qualitative) physical considerations in order to explain the results
of the simulations.


\section{The Einstein-Klein-Gordon system}
\label{equations}
The dynamics of a complex scalar field in a curved spacetime is
described by the following Lagrangian density (adopting
geometrical units, i.e. $G=c=1$) \cite{das}
\begin{eqnarray}\label{Lagrangian}
  {\cal L} = - \frac{1}{16 \pi} R
   + \frac{1}{2} \left[ g^{ab} \partial_a \bar \phi \partial_b \phi
   + V\left( \left|\phi\right|^2\right) \right]\, .
\end{eqnarray}
Here $R$ is the Ricci scalar, $g_{ab}$ is the spacetime metric,
$\phi$ is the scalar field,  $\bar \phi$ its complex conjugate,
and $V(|\phi|^2)$ a  potential depending only on $|\phi|^2$.
Throughout this paper Roman letters at the beginning of the
alphabet $a,b,c,..$ denote spacetime indices ranging from 0 to 3,
while letters near the middle $i,j,k,..$ range from 1 to 3,
denoting spatial indices.
This Lagrangian gives rise to the equations determining the
evolution of the metric (Einstein equations) and those governing
the scalar field behavior (Klein-Gordon Equations).

\subsection{Einstein Equations}
The variation of the action associated with the Lagrangian (\ref{Lagrangian}) with respect
to the metric $g_{ab}$ leads to the Einstein equations 
\begin{eqnarray}\label{EE1}
   R_{ab} - \frac{R}{2} g_{ab} = 8 \pi T_{ab},
\end{eqnarray}
where $R_{ab}$ is the Ricci tensor and $T_{ab}$ is
the stress-energy tensor given by
\begin{eqnarray}\label{stress-energy}
   T_{ab} &=& \frac{1}{2} \left[\partial_a \bar \phi \partial_b \phi
                        + \partial_a \phi~ \partial_b \bar \phi \right]
   \nonumber \\
    &-& \frac{1}{2} g_{ab} \left[g^{cd}\partial_c \bar \phi  \partial_d \phi
        + V\left(|\phi|^2\right)\right].
\end{eqnarray}
The Einstein equations form a system of $10$ non-linear partial
differential equations for the spacetime metric components
$g_{ab}$. A convenient way to express the equations can be obtained
by the identity
\begin{eqnarray}
R_{ab} &=& -\frac{1}{2} g^{cd} \partial_c \partial_d g_{ab} +
             \partial_{(a} \Gamma_{b)} - \Gamma_{cab} \Gamma^c \nonumber \\
       & & + g^{cd} g^{ef} \left( \partial_e g_{ca} \partial_f g_{db}
           - \Gamma_{ace} \Gamma_{bdf}  \right) \, .\label{ricci}
\end{eqnarray}
where we have introduced the quantities
\begin{eqnarray}
\Gamma_a &\equiv& g^{bc} \Gamma_{abc} \\
\Gamma_{abc} &=& \frac{1}{2} \left ( \partial_b g_{ac} + \partial_c g_{ab} - \partial_a g_{bc} \right ).
\end{eqnarray}

To this point we have considered arbitrary coordinates $x^a$ that label spacetime
points. In order to define an initial value problem for the Einstein equations, a
foliation defined by the hypersurfaces with $x^0\equiv t = const$ is introduced with
normalized normal vector $n_a = - \nabla_a t / ||\nabla_a t ||$, where $\nabla_a$ is
the covariant derivative associated to $g_{ab}$. 
The harmonic formulation of Einstein equations exploits the fact that the
coordinates $x^a$ can be chosen satisfying the
harmonic $\nabla^c \nabla_c x^a = 0$~\cite{Choquet55} or
their generalized version~\cite{Frie85,Garfinkle}
\begin{equation}
\nabla^c \nabla_c x^a = H^a(t,x^i) \label{harmonic1} \, ,
\end{equation}
where the functions $H^a$ amount to prescribing coordinate conditions. These expressions
can be combined to re-express Einstein equations (\ref{EE1}) in their generalized harmonic form,
as~\cite{Frie85,LSKOR06}
\begin{eqnarray}\label{dedonder1}
 && g^{cd} \partial_{cd} ~g_{ab}
 + \partial_{a} H_{b} + \partial_{b} H_{a}  =
 - 16~\pi~\left(T_{ab} - \frac{T}{2}~g_{ab}\right) \nonumber \\
 && + 2~\Gamma_{cab}H^c
  + 2~g^{cd}g^{ef}\left[~\partial_e g_{ac}~ \partial_f g_{bd}
 - \Gamma_{ace}~\Gamma_{bdf}~\right]
\end{eqnarray}
Notice that the partial derivatives of $H^a$ do not belong to the principal part
of the system since they are prescribed spacetime functions. The principal
part of (\ref{dedonder1}) is a well posed
system of decoupled wave equations for the 10 metric components.
It is important to stress that equations (\ref{dedonder1}) with the
generalized harmonic condition (\ref{harmonic1}) constitute
an overdetermined system since $\Gamma_a$ can
be obtained either from the choice of coordinates
(\ref{harmonic1}) or from the evolution of the metric itself
(\ref{dedonder1}). 
In the free-evolution approach, equation (\ref{dedonder1}) is
adopted to update the metric, and 
equation (\ref{harmonic1}) is regarded as a
constraint on the full system. Possible constraint violations can be measured
by introducing a new four-vector~\footnote{Notice the factor of $2$ difference
between the formulation presented here and \cite{LSKOR06}. This is for
convenience as unnecessary factors of $2$ are hence removed}
\begin{equation}\label{harmonicZ}
    2~Z^a \equiv - \Gamma^{a} - H^a(t,x^i)~~.  \end{equation}
Clearly, the physical solutions will be those satisfying the algebraic
condition $Z_a=0$ throughout the spacetime; we will refer to the $Z_a$ quantities
as the Z-constraints from now on. Note that added flexibility,
useful for numerical purposes, is attained by adding these
constraints to the equations. In particular,
since the addition of these Z-constraints to the evolution system can change
the stability of the solutions against perturbations off the constraint surface
one can exploit this fact to one's advantage. Indeed, suitable terms can be added to the
equations in order to construct an attractor for the physical
solutions in certain spacetimes, in such a way that small Z-constraint
violations will be damped during the evolution \cite{GCHM05}. 
With these damping terms, equation (\ref{dedonder1}) can be written as
\begin{eqnarray}\label{dedonder2}
 && g^{cd} \partial_{cd} ~g_{ab}
 + \partial_{a} H_{b} + \partial_{b} H_{a}  = \\
 && + 2~\Gamma_{cab}H^c
  + 2~g^{cd}g^{ef}\left[~\partial_e g_{ac}~ \partial_f g_{bd}
 - \Gamma_{ace}~\Gamma_{bdf}~\right] \nonumber \\
 && - 2~\sigma_0~\left[n_a Z_b + n_b Z_a - g_{a b} n^c Z_c \right]
 - 16~\pi~\left( T_{ab} - \frac{T}{2}~g_{ab}\right) \nonumber
\end{eqnarray}
where we have introduced $\sigma_0$, which is a
free parameter that controls the damping of the Z-constraints (\ref{harmonicZ}).

\subsection{Klein-Gordon Equations}
As mentioned, the variation of the Lagrangian (\ref{Lagrangian})
with respect to the scalar field $\phi$ leads to the Klein-Gordon (KG) equations
\begin{equation}\label{KG1}
  \Box \phi = \frac{d V}{d |\phi|^2} \phi \, ,
\end{equation}
where the box $\Box = g^{ab} \nabla_a \nabla_b$ stands here for
the wave operator on a curved background. For concreteness,
from now on we will consider the free field case where the potential takes the form
\begin{equation}\label{potential}
  V( |\phi|^2 ) = m^2~ |\phi|^2 \, ,
\end{equation}
with $m$ a parameter that can be identified with the bare
mass of the field theory, although it has units of inverse
length. For our simulations we have fixed its value to unity.
The potential (\ref{potential}) leads to the so-called
\textit{miniboson stars}, because achievable stable configurations have small masses. 
More general terms can be included, such as the $\lambda~|\phi|^4$ self-interaction term 
introduced in \cite{colpi}, leading to heavier \textit{boson stars} which have masses and 
sizes more relevant to astrophysical applications when considering their compactification
ratio. In the present work we restrict to
the $\lambda=0$ case, deferring to a future work the study of the effect of this self-interacting term.

\subsection{Reduction of the complete system to first order}
The second order Einstein-Klein-Gordon (EKG) equations (\ref{dedonder2})-(\ref{KG1}),
can be employed to obtain the future evolution of the spacetime if  $\{g_{ab}, \phi,
g_{ab,t}, \phi_{,t} \}$ are provided at an initial hypersurface.
For a numerical implementation, it is also useful to further reduce the system to fully
first order to take advantage of several numerical techniques devised to exploit 
stability theorems for these kind of systems~\cite{gustaffsonkreissoliger,olsson,strand,tadmor}.
Consequently, these techniques provide a clean path to discretize the system in a robust 
manner~\cite{SBP0,SBP1,SBP2}. Thus we begin by 
reducing these equations to first order form.

The reduction in time is achieved by introducing
new independent variables related to the time derivatives
of the fields. Following a similar route as  in \cite{LSKOR06}, we define
\begin{eqnarray}\label{time1}
  Q_{ab} \equiv -n^c~ \partial_c g_{ab}~,~~
  \Pi \equiv -n^c~ \partial_c \phi.
\end{eqnarray}
The evolution equations for $Q_{ab}$ and $\Pi$ are now given by the EKG
equations (\ref{dedonder2},\ref{KG1}), while the evolution equations
for $g_{ab}$ and $\phi$ are simply given by their definition (\ref{time1}).

The reduction to first order in space is made by introducing new independent
variables encoding the first space derivatives as
\begin{eqnarray}\label{space1}
  D_{iab} \equiv \partial_i g_{ab}~,~~
  \Phi_i \equiv \partial_i \phi.
\end{eqnarray}
The equations of motion for these quantities are obtained by applying the time
derivative to their definition (\ref{space1}) and imposing that
 time and spatial derivatives commute.
Notice that in this step, just as with the Z-constraints, one encounters
an overdetermined system since spatial derivatives can be
obtained either by deriving the fields or from the
evolution equation of the new quantities (\ref{space1}).
This fact can be re-interpreted as adding new constraints to the system
\begin{equation}\label{constraint1}
  {\cal C}_{iab} = \partial_i g_{ab} - D_{iab} = 0~,~~
  {\cal C}_i = \partial_i \phi - \Phi_i = 0~,
\end{equation}
which must be satisfied for a consistent solution.
An analogous situation is encountered with the conditions
\begin{eqnarray}\label{constraint2}
  && {\cal C}_{ijab} = \partial_i D_{jab} - \partial_j D_{iab} = 0 ~,~~ \nonumber \\
  && {\cal C}_{ij} = \partial_i \Phi_j - \partial_j \Phi_i = 0
\end{eqnarray}
resulting from the commutativity of the second spatial derivatives. 
Henceforth we will refer to the $\{{\cal C}_i,{\cal C}_{ij},
{\cal C}_{iab},{\cal C}_{ijab}\}$ quantities
as the C-constraints. At this point the resulting first order system is
described by the  evolution equations for the array of
fields $\{ g_{ab}, Q_{ab}, D_{iab},\phi,\Pi,\Phi_i \}$ together with
the Z-constraints (\ref{harmonicZ}) and C-constraints (\ref{constraint1},\ref{constraint2}).

Additionally, as with the Z-constraints, 
the mathematical properties of the evolution system can be changed
--when the constraints are not exactly satisfied-- 
by judiciously adding the C-constraints to the right-hand-side of the equations.
For instance, in \cite{LSKOR06} these constraints are incorporated in the
equations in such a way that the physical solutions (i.e., those
satisfying ${\cal C}_i = {\cal C}_{ij} = {\cal C}_{iab} = {\cal C}_{ijab} =0$) are an
attractor in certain spacetimes. This means that small C-constraint
violations will also be damped during the evolution. We follow the same
approach here for the full first order reduction of the Einstein
Klein-Gordon equations by writing them as:
\begin{eqnarray}\label{EKG1order}
  \partial_t g_{ab} &=& \beta^k~D_{kab} - \alpha~Q_{ab} \\
  \partial_t Q_{ab} &=& \beta^k~\partial_k Q_{ab}
  - \alpha \gamma^{ij} \partial_i D_{jab}
  - \sigma_1 \beta^k \partial_k g_{ab} \\
  &-& \alpha~ \partial_a H_b - \alpha~ \partial_b H_a +
  2~\alpha~ \Gamma_{cab}~ H^c \nonumber \\
  &+& 2\, \alpha\, g^{cd}~(\gamma^{ij} D_{ica} D_{jdb} - Q_{ca} Q_{db}
                   - g^{ef} \Gamma_{ace} \Gamma_{bdf}) \nonumber \\
  &-& \frac{\alpha}{2} n^c n^d Q_{cd} Q_{ab}
  - \alpha~\gamma^{ij} D_{iab} Q_{jc} n^c
  + \sigma_1 \beta^k D_{kab} \nonumber \\
  &-& 16 \pi \alpha~(T_{ab} - \frac{g_{ab}}{2} T) \nonumber \\
  &-& 2~\alpha~\sigma_0~[n_a Z_b + n_b Z_a - g_{a b} n^c Z_c ]  \nonumber  \\
   \partial_t D_{iab} &=& \beta^k \partial_k D_{iab}
  - \alpha~\partial_i Q_{ab}
  + \alpha~\sigma_1 \partial_i g_{ab}   \\
   &+& \frac{\alpha}{2} n^c n^d D_{icd} Q_{ab}
  + \alpha~\gamma^{jk} n^c D_{ijc} D_{kab}
  - \alpha~\sigma_1 D_{iab} \nonumber \\
   \partial_t \phi &=& \beta^k \Phi_k - \alpha~\Pi \\
   \partial_t \Pi &=& \beta^k \partial_k \Pi
  - \alpha~\gamma^{ij} \partial_i \Phi_j
  - \sigma_1 \beta^k \partial_k \phi  \\
  &-& \frac{\alpha}{2} \Pi~n^c n^d Q_{cd}
  - \alpha~\gamma^{ij}~\Phi_i n^c Q_{jc} \nonumber \\
  &+& \alpha~\Gamma_c~\Phi^c + \alpha~m^2~\phi
  + \sigma_1~\beta^k~\Phi_k \nonumber \\
   \partial_t \Phi_i &=& \beta^k \partial_k \Phi_i
  - \alpha~\partial_i \Pi + \alpha~\sigma_1 \partial_i \phi   
\label{lasteqn_EKG1order} \\
  &+& \frac{\alpha}{2}~\Pi~n^c n^d D_{icd}
  + \alpha~\gamma^{jk} \Phi_k n^c D_{ijc}
  - \alpha~\sigma_1 \Phi_i \nonumber
\end{eqnarray}
where $\sigma_1$ is another free parameter that controls the
damping of the first order constraints (\ref{constraint1}-\ref{constraint2}).
Additionally, we have introduced the familiar lapse $\alpha$ and shift $\beta^i$ functions
for convenience. These are defined through the relations $\beta^i = \gamma^{ij} \beta_j$ 
($\gamma_{ij}$ being the spatial projection of $g_{ab}$ satisfying $\gamma^{il} \gamma_{lj} = \delta^i_j$, $g_{ti} = \beta_i$) and $g^{tt} = \alpha^{-2}$.
Equations (\ref{EKG1order}-\ref{lasteqn_EKG1order}) constitute
the final set of equations being integrated in our implementation.\\

\subsection{Analysis quantities}
One way to identify the position of the boson star, which in the
fully dynamical might prove difficult, is to use the energy density
$\rho \equiv n^a\,n^b\,T_{ab}$. We have used the maximum of $\rho$, at 
each temporal slice to represent the center of each star-like configuration.

Another useful quantity, which is of special interest in the
case of the boson-antiboson collision, is the Noether charge of the theory.
Due to the U(1) symmetry of the stress-energy tensor, there is a conserved quantity
which can be computed by
\begin{equation}
  N = \int J^0~ dx^3 ~~,~~
  J^0 = \frac{i}{2} \sqrt{-g} g^{0 \nu} \left[\bar \phi ~\partial_{\nu} \phi-
  \phi~\partial_{\nu} \bar \phi \right].
\end{equation}
We employ the Noether density $J^0$ to give
additional evidence as to the location of the boson star.  
As discussed in~\cite{ruffini}, this quantity can be associated 
with the number of bosonic particles.

The mass of the stars is computed by means of the ADM mass, which is defined as:
\begin{equation}\label{defmass}
  M = \frac{1}{16\pi} \lim_{r \rightarrow \infty}
  \int g^{ij}~\left[\partial_j g_{ik} - \partial_k g_{ij}\right]~{\cal N}^k~dS
\end{equation}
where ${\cal N}^k$ stands here for the unit outward normal to the sphere.

The gravitational radiation is described asymptotically --when the correct tetrad
and coordinates are adopted--, by the Newman-Penrose (spin-weighted) $\Psi_4$ scalar.
This quantity is defined as
\begin{equation}\label{defpsi4}
   \Psi_4 = C_{abcd}~ {\mathtt n}^a~ {\mathtt {\bar m}}^b~{\mathtt n}^c~ {\mathtt {\bar m}}^d~~,
\end{equation}
with $C_{abcd}$ being the Weyl tensor and where ${\mathtt n}^a$ denotes the incoming null normal to
the extraction worldtube $\Gamma$ while ${\mathtt {\bar m}}^a$ is the complex conjugate of the null vector
tangent to $\Gamma$ at a constant time slice.

To analyze the structure of the radiated waveforms it is convenient
to  decompose the signal in $-2$ spin weighted spherical harmonics as,
\begin{equation}\label{rpsi4}
  r~\Psi_4 = \sum_{l,m} C_{l,m} {}^{-2} Y_{l,m}
\end{equation}
where the factor $r$ is included to better capture the $1/r$ leading order behavior of $\Psi_4$.
For head-on collisions of non-spinning objects, one could take advantage of the natural axis of symmetry
defined by the line joining the centers of the objects.  In such a case, 
the (spin weighted) spherical harmonics ${}^{s} Y_{l,m}$ can be defined such that the radiation
produced in the axisymmetric configuration would display essentially only $m=0$ modes.
However, as we are interested in more general cases also, we adopt spherical
harmonics defined with respect to an axis that can be regarded as orthogonal with respect to
the orbital plane, which in our present case is given by the $z$ axis. In this case, the radiation
extracted would have non-trivial components for $m\neq0$.

We also focus in integral quantities that are independent of the specific
basis of the spherical harmonics, such as the radiated energy. In terms of $\Psi_4$, we can write
\begin{equation}\label{dEdt1}
  \frac{dE}{dt} = \frac{r_{ext}^2}{4\pi} \int_{\Omega} \left[~\left| \int^t_{-\infty} \Psi_4 (t')~ dt' \right|~
  \left| \int^t_{-\infty} {\bar \Psi}_4 (t')~ dt' \right|~ \right]~ d{\Omega} ~,
\end{equation}
where ${\bar \Psi}_4$ denotes the complex conjugate of $\Psi_4$, and $r_{ext}$ is the extraction radius. 
Using both the decomposition (\ref{rpsi4}) and the orthonormalization of the spherical harmonics,
this expression can also be written as
\begin{equation}\label{dEdt2}
  \frac{dE}{dt} = \frac{1}{4\pi} \sum_{l,m} | D_{l,m} |^2
  ~~,~~ D_{l,m} = \int^t_{-\infty} C_{l,m}(t')~ dt'.
\end{equation}
Finally, we notice that we will assume the resulting coordinates in the neighborhood
of the extraction surfaces satisfy, to a reasonable level, conditions ensuring
$\Psi_4$ is essentially free of spurious effects (aside from those arising from the
extraction at finite distances)~\cite{lehnermoreschi,baumgarteetal}. Detailed studies of 
the possible influences of these in more general settings will be presented elsewhere~\cite{lsueffects}.


\section{Implementation Issues}
\label{numerics}
The first order symmetric hyperbolic system (\ref{EKG1order}-\ref{lasteqn_EKG1order})
is implemented following techniques devised to take advantage of several useful theorems 
which guarantee the stable implementation of linear hyperbolic systems.
In general these techniques involve:
(i) discrete derivative operators satisfying summation by parts (SBP), which allow
defining a semidiscrete norm and an energy estimate 
for the solution \cite{strand,gustaffsonkreissoliger},
(ii)  a method of lines employing a third order Runge-Kutta operator for the
time integration that preserves the energy estimate in the fully discrete case
\cite{tadmor}, and
(iii) maximally dissipative boundary conditions that keep this energy bounded
in time \cite{olsson} in presence of boundaries. 
The combination of techniques (i-iii) guarantee that linear problems are implemented
stably and thus provides a direct route to a robust implementation of
generic first order hyperbolic systems. Additionally, we introduce a Kreiss-Oliger type 
dissipative operator (suitably modified at/near boundary points so as not to spoil the 
energy estimate \cite{SBP0,SBP1}). The introduction
of dissipation allows one to control the high frequency modes of the solution which are
always poorly described by a finite difference approach. In what follows we provide further details
of the particular application of these techniques in the present work.

\subsection{Discrete operators and time evolution}

We have adopted second order derivative
operators satisfying SBP and defined a discrete energy essentially as the $L_2$ norm of the solution,
that is
\begin{equation}\label{energy}
   E =  \sum_{A} || U_A || \equiv  \sum_{A} \sum_{I,J,K} U_A(I,J,K)^2
\end{equation}
where $U_A = \{ g_{ab}, Q_{ab}, D_{iab}, \phi, \Pi, \Phi_i \}$ is an array with all the 
evolved variables and $U_A(I,J,K)$ is the value of $U_A$ at the grid point $(I,J,K)$.
The energy estimate is obtained when the discrete operators reduce to the standard
three-point centered derivative operator at interior points while the standard
(first order) two-point sideways stencil at boundary points . The particular form for the
Kreiss-Oliger fourth order dissipation operator is described in \cite{SBP0,SBP1} as is 
the third order Runge-Kutta scheme employed to advance the solution in time.

We additionally employ the norm of the Z-constraints,
\begin{equation}\label{Zenergy}
  || Z || =  \sum_{a=0}^{3} || Z_a ||
\end{equation}    
to monitor the deviation of the numerical solution from the physical one.

\subsection{Characteristic structure}
As mentioned, we employ maximally dissipative boundary conditions. These conditions
rely on knowing the characteristic mode structure of the principal part of the system.
In essence, these conditions bound the amount of ``energy'' the incoming modes introduce at the boundaries
to remain below what the outgoing modes carry away from the computational domain. Schematically,
this reduces to adopting conditions satisfying $w^{-} = L w^{+}$ with $w^{-},w^{+}$ denoting
incoming/outgoing modes at a given boundary, and $L$ a bounded matrix \cite{gustaffsonkreissoliger}.
As a result, the norm~(\ref{energy}) of the solution remains bounded in the linear regime, which is key to ensure the well
posednes of the IBVP (for details see \cite{gustaffsonkreissoliger,olsson,tadmor}). In our particular 
case, we choose $L=0$ thus setting the incoming modes to zero.
The characteristic decomposition (eigenmodes $w$ with velocity $v$)
for the EKG system (\ref{EKG1order}-\ref{lasteqn_EKG1order}) is given by
\begin{eqnarray}
  {w_{ab}}^0 &\equiv& g_{ab} ~~~~~~~~~~~~~~~~~~~~~~~~~~~~,~  v^0 = 0 \\
  {w_{iab}}^\beta &\equiv& D_{iab} - {\cal N}^k~D_{kab}~{\cal N}_i 
  ~~~~~~~,~  v^\beta = -{\cal N}_k~\beta^k \nonumber \\
  {w_{ab}}^{(\pm)} &\equiv& Q_{ab} - \sigma_2~ g_{ab} \pm {\cal N}^k~D_{kab}
  ~,~  v^{\pm} = -{\cal N}_k~\beta^k \pm \alpha \nonumber
\end{eqnarray}
with ${\cal N}^k$ the outgoing normal vector to the boundary surface and the supra-indices $\{ 0,\beta,\pm\}$
denote modes propagating with the different speeds $\{ v^0,v^{\beta},v^{\pm}\}$.
Thus the incoming modes are given by $w_{ab}^-$ and ${w_{iab}}^\beta$ as long as ${\cal N}_k~\beta^k < 0$.
To apply
the maximally dissipative conditions at the numerical level, we follow the prescription described in \cite{olsson,SBP0}
and simply enforce
\begin{eqnarray}
   \partial_t {w_{ab}}^{(-)} &=& 0,\ \  \\
   \partial_t w_{iab}^\beta  &=& 0, \ \  \ \ {\rm if} \ \ \ \ \ {\cal N}_k~\beta^k < 0.
\end{eqnarray}


\subsection{Mesh Refinement}
Adaptive mesh refinement~(AMR) provides increased resolution where and when needed,
and therefore becomes an essential tool in resolving a wide range of dynamical scales, both
spatial and temporal. As described by Berger and Oliger~\cite{berger2}, a given numerical
grid can be further refined based on dynamical measures of the numerical error by
the creation of refined grids which are evolved in sync with the coarse grid.
We have built our code taking advantage of the computational
infrastructure called HAD (an outgrowth of the code presented in \cite{had1,had2}).
HAD implements a (slight) modification of the standard
Berger-Oliger strategy that guarantees preserving
the stability properties of the unigrid implementation and significantly
reduces spurious reflections at interface boundaries~\cite{amr}.
In HAD creation or destruction of finer grids is automated so that points which display
an error above some threshold are ``covered'' by a finer grid.  The error associated with
each point is computed via a self-shadow hierarchy~\cite{pretoriusphd}.
In this technique, the error is computed on any given grid by subtracting the solution on
that grid with that on the next coarser level. Because these solutions evolve with
different resolutions, their difference represents a measure of the local truncation error
without recourse to more complicated shadow schemes. In particular, at any given point,
the $L_2$ norm of the difference between the solutions over all evolved fields 
is taken as the truncation error estimate.
Although we have performed binary boson star runs employing AMR, 
in this work we take only partial advantage of the mesh refinement capabilities of HAD.
One reason for this is that we want to compare significantly different scenarios and we find
it useful to ensure similar discretizations are employed across the cases 
considered.  Our procedure is to let the infrastructure define the grid structures at the initial
time and keep it fixed throughout the simulations. The truncation error value employed
by the self-shadow hierarchy to define the grids is chosen such that the original stars
are contained within the finest grid and that this grid covers the whole region in between.
Fixing the grid structure at the initial time reduces some of the overhead and memory required by dynamic regridding.


\section{Simulations and Results}
\label{results}

In this section we describe the simulations we have performed both
to check the correctness of the implementation and investigate interesting
aspects of boson star head-on collisions. We begin in~\ref{single_boson},
by considering a single isolated boson star configuration and verify the
code is able to reproduce known features of the solution.
The evolution of the star is performed employing two different coordinate
systems, a static one and the harmonic one. The former coordinate system
is the natural one employed in constructing a single static boson star solution,
and so the evolution with it should be trivial.
The latter one does not conform to the ansatz employed to obtain the solution
and some non-trivial dynamics can be expected.

In Section~\ref{headon} we consider different binary systems described
by boson stars with zero initial momentum. Initial data for these systems is
obtained by a simple superposition of a single boson star with another which is
identical to it up to a possible phase or reflection of its natural frequency.
The stars are placed sufficiently far from each other to ensure the constraints
 are satisfied to a degree consistent with the value obtained in the single boson star case.

We have performed three different types of head-on collisions: collisions
of equal boson stars~\ref{equal_mass_merger}, collisions of boson stars
whose phase are in opposition~\ref{unequal_phase_merger} and finally
collisions of  boson stars with antiboson stars~\ref{boson_antiboson}
(these differences are explained in detail below).

For all of the simulations presented here we have adopted damping parameters 
$\sigma_0=\sigma_1=1$, a Courant factor $\Delta t/\Delta x=0.25$ and
Kreiss-Oliger dissipation parameterized by $0.03$ (as described in~\cite{SBP0,SBP1}). The simulations have been performed on
64 to 128 processors employing 3 refinement levels, placed as defined
by the shadow hierarchy.


\subsection{The single boson star}
\label{single_boson}
As described in the introduction, a boson star is a self gravitating 
configuration of scalar field with Lagrangian given by~(\ref{Lagrangian}). 
In a particular coordinate system, one assumes a particular form for
the spherically symmetric scalar field
\begin{equation}
\label{oscillatory_ansatz}
\phi = \phi_0(r)~e^{-i \omega t}\, .
\end{equation} 
Therefore the field values at any given point oscillate 
with a constant frequency $\omega$. Details of the calculation of 
$\phi_0(r)$ are presented in Appendix~\ref{initial_data}. The 
geometry, on the other hand, remains static in these same coordinates.
The resulting spacetime provides a reasonable model for that resulting from
a compact object in equilibrium.

We have carried out simulations of single boson stars under two coordinate
choices. In~\ref{static_evolution} we summarize the results in the coordinate
system on which the solution is static and the results in the
harmonic gauge is presented in~\ref{harmonic_coords}. The simulations presented
for both cases employ a grid structure consisting of three levels of refinement.
The finest level covers the radius $R_{95}$ of the star 
(defined as region of space containing approximately 95\% of the star's mass) and 
the other boxes surrounding it with half the resolution each respectively. 
For all these runs the star is located 
at $x^i=0$ ($i=1..3$) and the computational domain is given by  $x^i \in [-144,144]$.

To obtain a measure of the convergence rate of the solution, we have performed 
simulations with different resolutions for the boxes. The finest grid of each
resolution considered has $\Delta x \in \{0.375, 0.25, 0.156\}$ 
for the static coordinates and $\Delta x \in \{0.5, 0.375, 0.25\}$ for the harmonic ones.
The others grids, as mentioned, have a gradual 2:1 ratio in their discretization length.

Below we detail certain convergence tests of the code about which we would like to make a general comment.
Residuals of constraints are expected
to approach zero with increases in resolution while field variables should approach a unique solution, and
this behavior is indeed what we find.
The rate of convergence can also be examined and compared to the expected approximation order of the
difference equations solved, and we found the expected second order convergence in the unigrid case.
However when dealing with a complicated grid hierarchy, precise measurements of this rate become
somewhat problematic, especially when the hierarchies produced vary with the truncation
error threshold chosen. Also, we should mention that more detailed convergence tests with
this code infrastructure applied to other problems
demonstrate the expected order of convergence~\cite{amr,matt_paper}.

\subsubsection{Single boson star in static coordinates}
\label{static_evolution}
In order to maintain the coordinates such that the spacetime remains static
the $H^a(t,x^i)$ functions in~(\ref{harmonic1}) need to been chosen in a suitable
way. Since we are considering a static situation, the condition on $H^a$ can be
read directly from the equation (\ref{harmonicZ})
by imposing the constraints (i.e., $Z^a = 0$) and the staticity of the spacetime (i.e., 
$\partial_t g_{ab}=0$). In a explicitly static scenario, this initial value of the $H^a(t=0,x^i)$
is preserved for all times.

\begin{figure}[t]
\begin{center}
\epsfxsize=8cm
\epsfbox{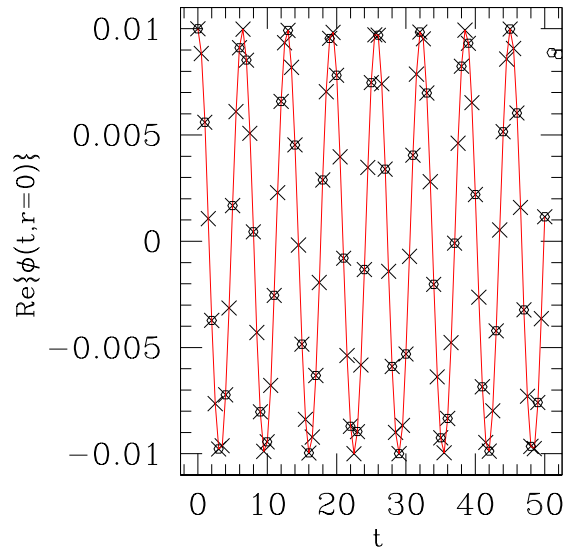}
\end{center}
\caption{ \textit{Single boson}. Phase oscillation of the real part of the scalar field at
the center of symmetry up to $t=50$ for the resolution $\Delta x =0.25$ (similar behavior was followed up to $t=500$). The solid line indicates the analytically
expected value, the crosses show the values read-off from the numerical solution within
the static coordinates and the circles show the same by using the harmonic coordinates.}\label{sphi}
\end{figure}

\begin{figure}[t]
\begin{center}
\epsfxsize=8cm
\epsfbox{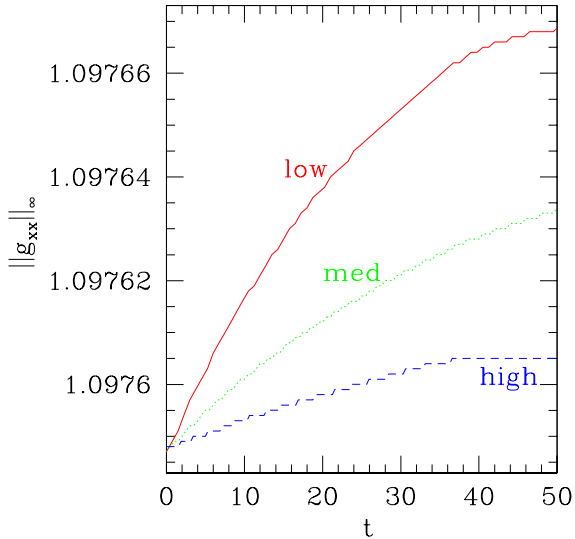}
\end{center}
\caption{ \textit{Single boson}. $L_\infty$ of $g_{xx}$ versus time for solution obtained employing the 
static 
coordinate condition. The figure displays the result of the evolution for three different base-resolutions 
$\Delta x = \{0.375, 0.25, 0.156\}$ (in addition, each numerical solution is obtained with three levels
of refinement).
The values obtained converge to a constant value as expected, and the maximum variation observed to $t=50$
is $\simeq 0.0073\%$ for the coarsest base-resolution while for the highest one is just $\simeq 0.0015\%$.
}\label{sg11}
\end{figure}

We begin with an initial data describing a star where the 
value of the scalar field defining it at the origin is  
$|\phi_0(r=0)|=0.01$. This star, which is obtained by solving 
the initial data problem with a high resolution one-dimensional code as described in the
Appendix~\ref{initial_data}, has an ADM mass of $M_{1D}=0.361$ and radius $R_{95}=19.6$, and
lies well inside the stable branch of solutions. 

The evolution of an isolated star displays the expected behavior. In particular, the
geometry is static and the behavior of the scalar field agrees with Eq.~(\ref{oscillatory_ansatz})
as demonstrated in Fig.~\ref{sphi} where the central value of the 
real part of the scalar field versus time describes an oscillatory behavior with a frequency
$\omega=0.96 \pm 0.03$. This is in excellent agreement with the frequency obtained with the
one-dimensional initial-data problem code $\omega_{1D}=0.976$.

\subsubsection{Single boson star in harmonic coordinates}
\label{harmonic_coords}
In this section we revisit the initial data of the previous one, but in this case
we do not adopt coordinate conditions that explicitly demonstrate 
the staticity of the spacetime. One goal of this exercise is to test the ability
of the harmonic-coordinate condition to adapt to the physical problem under consideration.
This will be important for the binary cases considered later, as that problem is certainly
not static. 

The implementation evolves the system for as long as the code was run without
any signs of instabilities. For this particular case we have evolved the system 
for more than $t=1500$. During this evolution, we have monitored the ADM mass which
we extract at $r_{ext}=100$ and obtain $M=0.364$ which decreases by less than $3\% \, / \,  1 \% $
for the coarsest/finest resolution considered ($\Delta x=0.5\, / \, 0.25$ for the innermost grid) 
by the time we stop the code. This illustrates the 
ability of the code to preserve this conserved quantity for long times. 

We have also checked that for this particular solution, both the harmonic and
the static coordinates agree at the center of symmetry. This implies that
the scalar field should have the same local oscillatory behavior~(\ref{oscillatory_ansatz})
at $r=0$ in both coordinate systems. This is indeed the case and is illustrated in 
Fig.~\ref{sphi}. In particular, the measured oscillation frequency is $\omega = 0.97 \pm 0.06$,
Additionally, we have checked that the absolute value of scalar field at the 
center of the star varies by at most $12\%\, / \, 2 \%$ for the coarsest/highest resolution considered.

\begin{figure}[t]
\begin{center}
\epsfxsize=8cm
\epsfbox{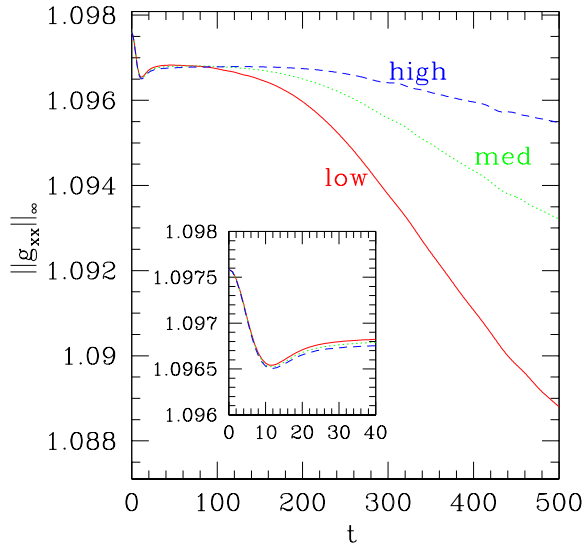}
\end{center}
\caption{\textit{Single boson}. $L_\infty$ of $g_{xx}$ versus time for solution obtained employing the harmonic coordinate condition. The figure displays the result of the evolution for three different base-resolutions
$\Delta x = \{0.25, 0.375, 0.50\}$ (in addition,
each numerical solution is obtained with three levels of refinement).
After a small transient behavior, the values obtained converge to a constant value. The maximum variation observed to $t=500$ is $\simeq 0.72\%$ for the coarsest base-resolution while $\simeq 0.11\%$  
for the highest one.
}\label{singleg11}
\end{figure}

Coordinate effects do arise however. As mentioned above the geometry should have  some non-trivial
coordinate-induced dynamics since we are not adopting a coordinate system on which the solution is
explicitly static.  This effect can be seen in  Fig.~\ref{singleg11} where the evolution of the maximum
of $g_{xx}$  as a function of coordinate time $t$ for 3 different resolutions is displayed. 
As evident in the figure, there is an initial transient variation of the metric value
which later approaches a constant value.

\begin{figure}[h]
\begin{center}
\epsfxsize=8cm
\epsfbox{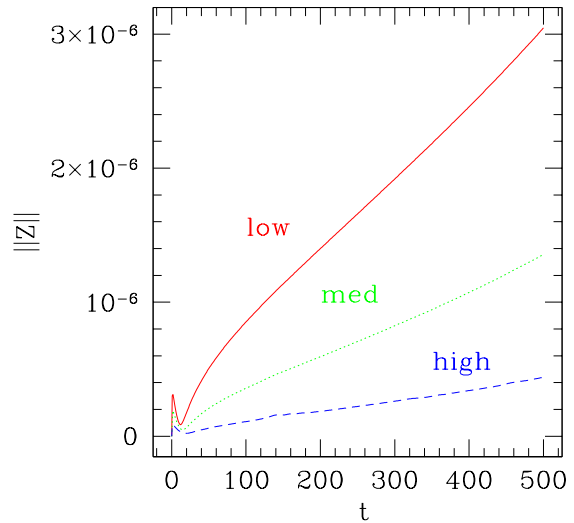}
\end{center}
\caption{\textit{Single boson}. Convergence of the $L_2$ norm of the Z-constraints residual 
(defined by equation~(\ref{Zenergy}))
versus time by using harmonic coordinates, for the same resolutions than in Fig. \ref{singleg11}.
As the resolution is increased the constraint violation is clearly reduced.}
\label{singleZconv}
\end{figure}

Finally, we monitor the constraints in Fig.~\ref{singleZconv}. The figure 
shows the $L_2$ norm~(\ref{Zenergy}) of the physical constraints (i.e., the Z-constraints~(\ref{harmonicZ})) and
its behavior when the resolution is improved, converging clearly to zero.
The measured convergence rate 
is  $2.9$ between the low-and-medium  and  $2.8$ between the medium-and-high resolutions. 

\subsection{Binary boson-stars head-on collisions}
\label{headon}
Initial data for the collision of two boson stars is obtained by a
simple superposition of the single boson solution in a way described below.
Additionally, in order to consider different configurations, we take advantage
of the following observations:\\

\indent $(i)$ The solution of the initial data problem is unaffected by a phase
difference in the ansatz assumed for the scalar field. Namely, $\phi = \phi_0(r) e^{i (w t + \theta)}$
gives rise to the same initial data for $\{g_{ab},\partial_t g_{ab}=0\}$.\\
\indent $(ii)$ if $\{g_{ab},\, \phi_0, \, \partial_t g_{ab}=0,\, \partial_t \phi =i\omega \phi_0 \}$  provides consistent
initial data, so does
$\{g_{ab},\phi_0, \partial_t g_{ab}=0, \partial_t \phi=-i\omega \phi_0 \}$. The difference is solely
in a frequency-reflection of the boson star and is known as an anti-boson star. \\

We will exploit these observations to consider what we will refer to
as a boson in phase opposition by taking $\theta=\pi$ in (i) and/or
an anti-boson in (ii).

The initial data we consider is schematically represented by considering the
following construction
\begin{eqnarray}\label{ID_equalcase}
  \phi &=& \phi^{(1)}(x-x_1,y,z) + \phi^{(2)}(x-x_2,y,z) \\
  \psi &=& \psi^{(1)}(x-x_1,y,z) + \psi^{(2)}(x-x_2,y,z) - 1 \nonumber \\
  \alpha &=& \alpha^{(1)}(x-x_1,y,z) + \alpha^{(2)}(x-x_2,y,z) -1 \nonumber
\end{eqnarray}
where $u^{(i)}$ denotes the corresponding field of the boson $i$, centered
at $(x_i,0,0)$ and the value of $\phi^{(i)}$ will be dictated by the type of
boson star considered, that is: a boson star, a boson star in phase opposition
or an anti-boson star. Notice that three fields (the scalar field $\phi$,
the conformal factor $\psi$ and the lapse $\alpha$) are enough to set an
initial data consistent with the EKG equations, as it is described in the appendix.
Under this approach we are making the assumption that the boson stars are described
with a single global scalar field $\phi$ instead of considering two sets of complex
scalar fields to represent each star. This choice fixes in a straightforward
way the interaction of the stars.
 
\begin{figure}[h]
\begin{center}
\epsfxsize=8cm
\epsfbox{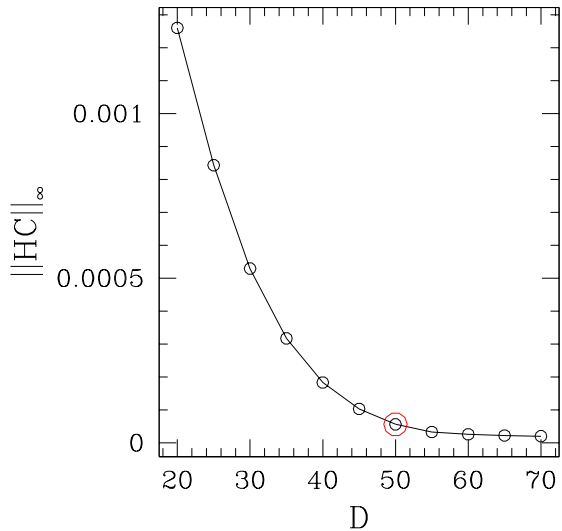}
\end{center}
\caption{\textit{Boson/Boson pair}. $L_{\infty}$ norm of the Hamiltonian constraint residual versus the initial (coordinate) separation of the boson stars. As the distance is increased the violation decreases
as expected since each star is defined so as to satisfy the constraints. We choose a value of $D=50$ which
lies close to the asymptotic behavior and is still manageable in terms of the boundaries.}\label{ID_conv}
\end{figure}

In order to consistently choose  the initial position of the boson stars we have measured, at
the initial time, the $L_{\infty}$ norm of the Hamiltonian constraint for different
coordinate separations of the centers $D$.  Since at sufficiently large distances the adopted data satisfies 
the constraints by construction, the constraint's behavior versus $D$ provides a measure of the
distance that must be adopted.  As illustrated in Fig.~\ref{ID_conv}, the error decreases
very rapidly with separation and we have chosen the value $D=50$ (which corresponds to $x_1=-x_2=25$) 
that is nearing the asymptotic behavior.
In addition, we will use for all the head-on collisions the same form of the scalar
function $\phi_0(r)$ in order to compare easily the different cases.

\subsubsection{Merging of the boson/boson pair}
\label{equal_mass_merger}

In this case, two identical boson star configurations are adopted to define the initial
data. Since the stars adopted have no initial velocity, their initial behavior is marked by
a slow approach towards each other as they feel their gravitational attraction. As the evolution
progresses however, the resulting behavior depends on the mass of the
single initial boson stars. Previous studies which concentrated on a particular set of initial
masses, always displayed a collision of stars giving rise to the formation of
a black hole \cite{balakrishna_phd}. More recently, the work of~\cite{lai,laipaper}, which
considered boosted stars with large kinetic energy, presented cases where
the stars appear to pass through each other in a sort of ``solitonic'' behavior.

\begin{figure}[h]
\begin{center}
\epsfxsize=8cm
\epsfbox{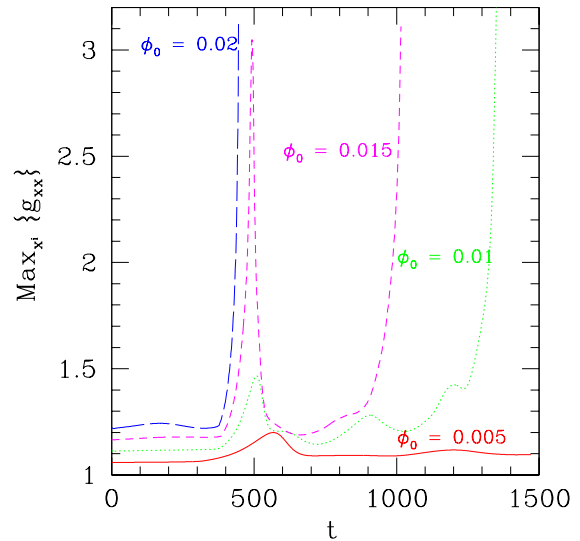}
\end{center}
\caption{\textit{Boson/Boson pair}. Geometry behavior for different values of central value of 
the scalar field that defines the stars. As the stars merge, the behavior of geometric variables
changes drastically for different values. Stars with initial central densities $>0.005$ 
seem to give rise to a black hole as illustrated by their considerable growth;
those below this value yield  much smoother behavior. This is illustrated in the figure
which displays the maximum of the component of the metric $g_{xx}$ versus time
for different central amplitudes of the scalar field $\{\phi_0=0.02,0.015,0.01, 0.05\}$ respectively.
These give rise to stars with to masses $\{M = 0.47,0.42,0.36,0.26\}$ respectively. }\label{maxg11}
\end{figure}

Here we want to study the situation in which there is a final regular object 
(that is, containing no singularities). To do so we begin with 
a broad parameter search along the masses of the 
boson stars searching for cases where the final compact
object does not collapse to a black hole. As shown in Fig.~\ref{maxg11}, 
for stars with masses around $M=0.26$ (corresponding to an amplitude of
$\phi_0(r=0)=0.005$) the final object appears to avoid black hole formation. This avoidance is indicated by
the geometric variables tending to a smooth slow oscillatory behavior for $M < 0.26$
while a marked violent behavior is displayed for larger values, together with a collapse
of the lapse function.

We thus focus on the particular case where each star has masses $M=0.26$, which is
approximately $40 \%$ of the maximum allowed mass on the stable branch for the potential
(\ref{potential}). The radius of the single star is around ${R_{95}}=27$, and
in this case we extend the computational domain to $x^i\in [-320,320]$.
As mentioned above, the refinement regions are rectangular boxes covering the centers of the stars 
and the span between them with a grid spacing at the region of the star of $\Delta x=0.50$ for the lowest
resolution and $\Delta x=0.375$ for the highest one. We compute the values
of the ADM mass and $\Psi_4$ at extraction surfaces
located at $r_{ext}= 140,~170$ and $200$, where the grid spacing is given by $\Delta x=4$ for
the lowest resolution and $\Delta x=3$ for the highest one (i.e., there are 3 levels
of refinement). Since the wavelength
of the observed radiation is of the order $\lambda \simeq 100$, the traveling wave is
well represented by the grid-structure at the extraction location.

\begin{figure}[h]
\begin{center}
\epsfxsize=8cm
\epsfbox{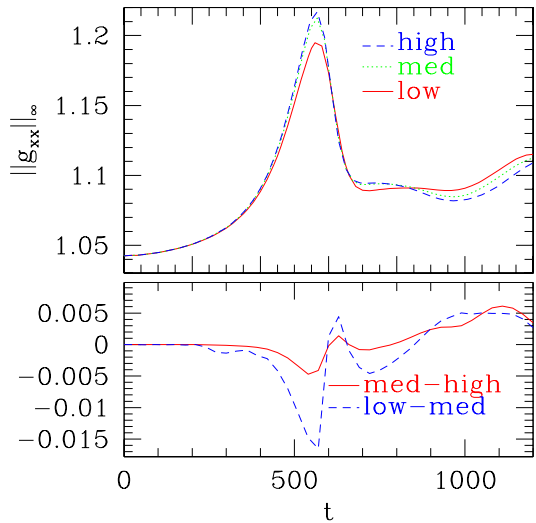} 
\end{center}
\caption{ \textit{Boson/Boson pair}. $L_{\infty} (g_{xx})$ as function of time for three
different resolutions and their differences. We can observe clearly that the merger
occurs at approximately $t\simeq 550$. The solution is qualitatively convergent during the
first part of the merger until $t\simeq 800$ when boundary effects negatively affect the convergence.
\label{2equal_maxg11_conv}}
\end{figure}

In an effort to ensure fidelity to the continuum problem, we monitor constraint residuals
and convergence of the metric variables for these evolutions.
Notice however that, again,
the grid structure for different resolutions differs and so our evolutions do not
lend themselves to a traditional convergence study. However we do see the expected
behavior, in that the constraint residuals decrease in regions where the resolutions differ and
the fields indicate convergence to a common value. This is illustrated 
in Fig.~\ref{2equal_maxg11_conv} where the maximum value of $g_{xx}$ versus time is displayed.
As evident in the plots, the values for different resolutions agree quite well until
a time of about $t\simeq800$ where accumulation of errors and boundary effects become apparent. 

\begin{figure}
\centering
\begin{tabular}{c}
\epsfig{file=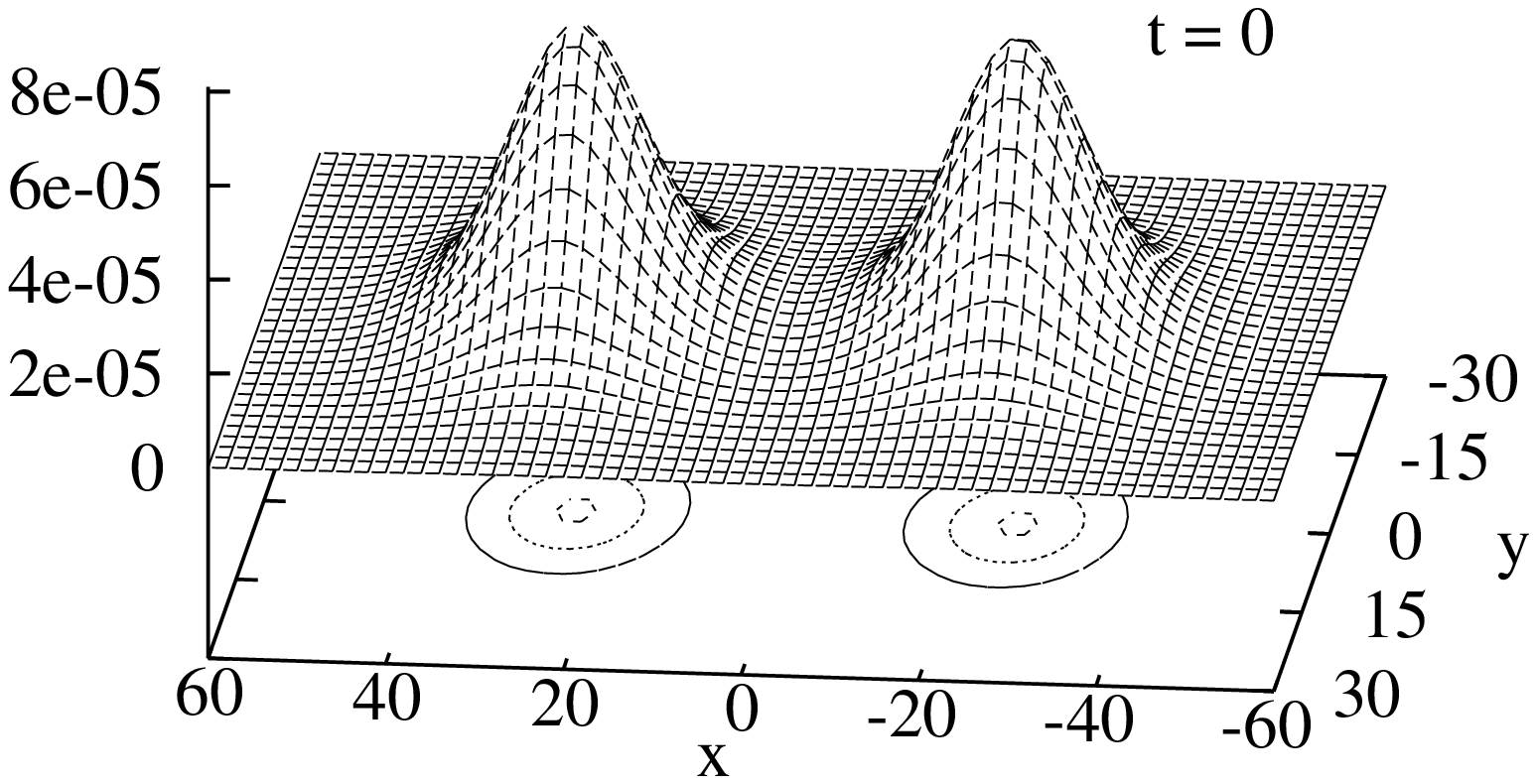,width=0.7\linewidth,clip=} \\
\epsfig{file=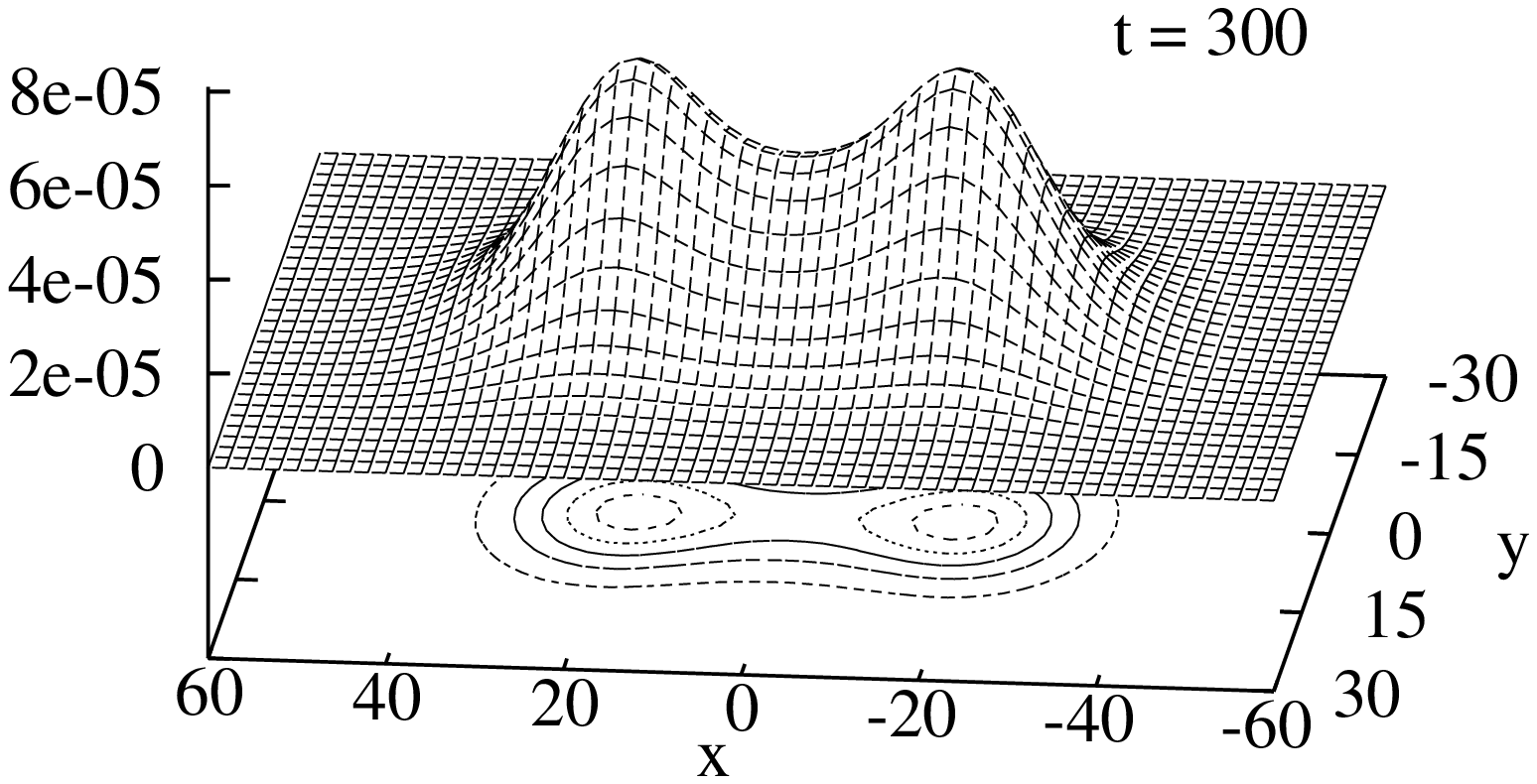,width=0.7\linewidth,clip=} \\
\epsfig{file=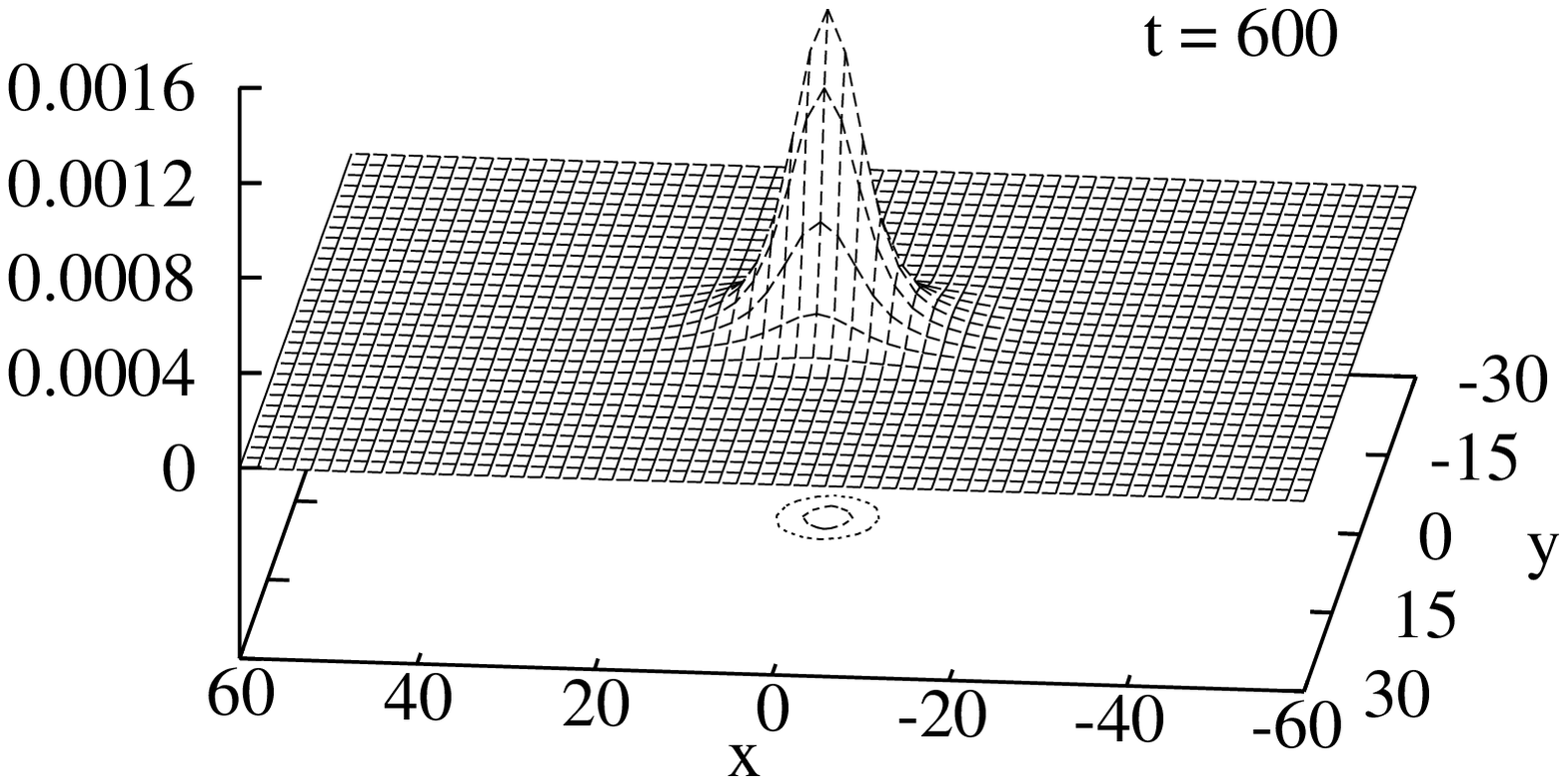,width=0.7\linewidth,clip=} \\
\epsfig{file=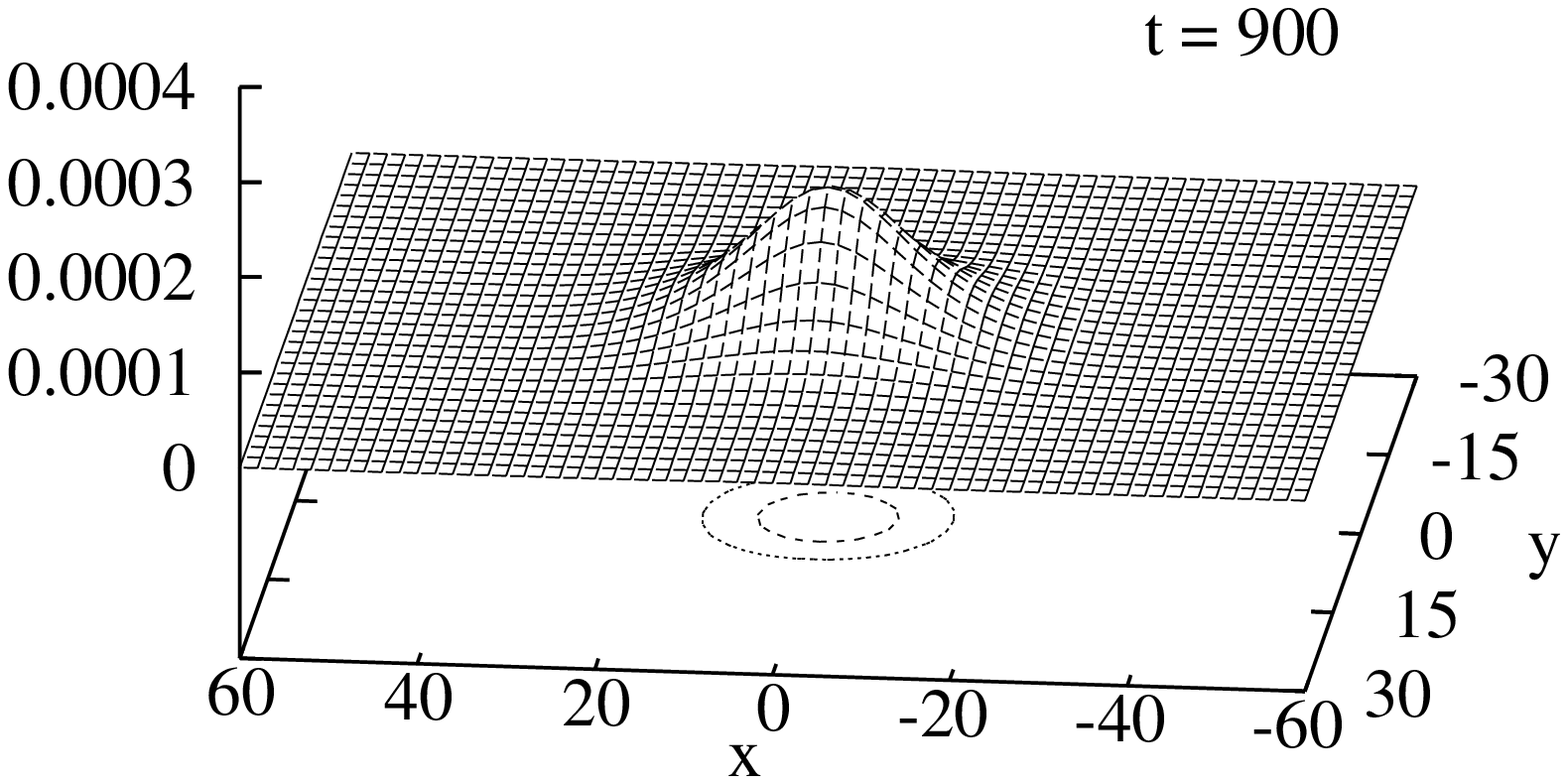,width=0.7\linewidth,clip=} \\
\epsfig{file=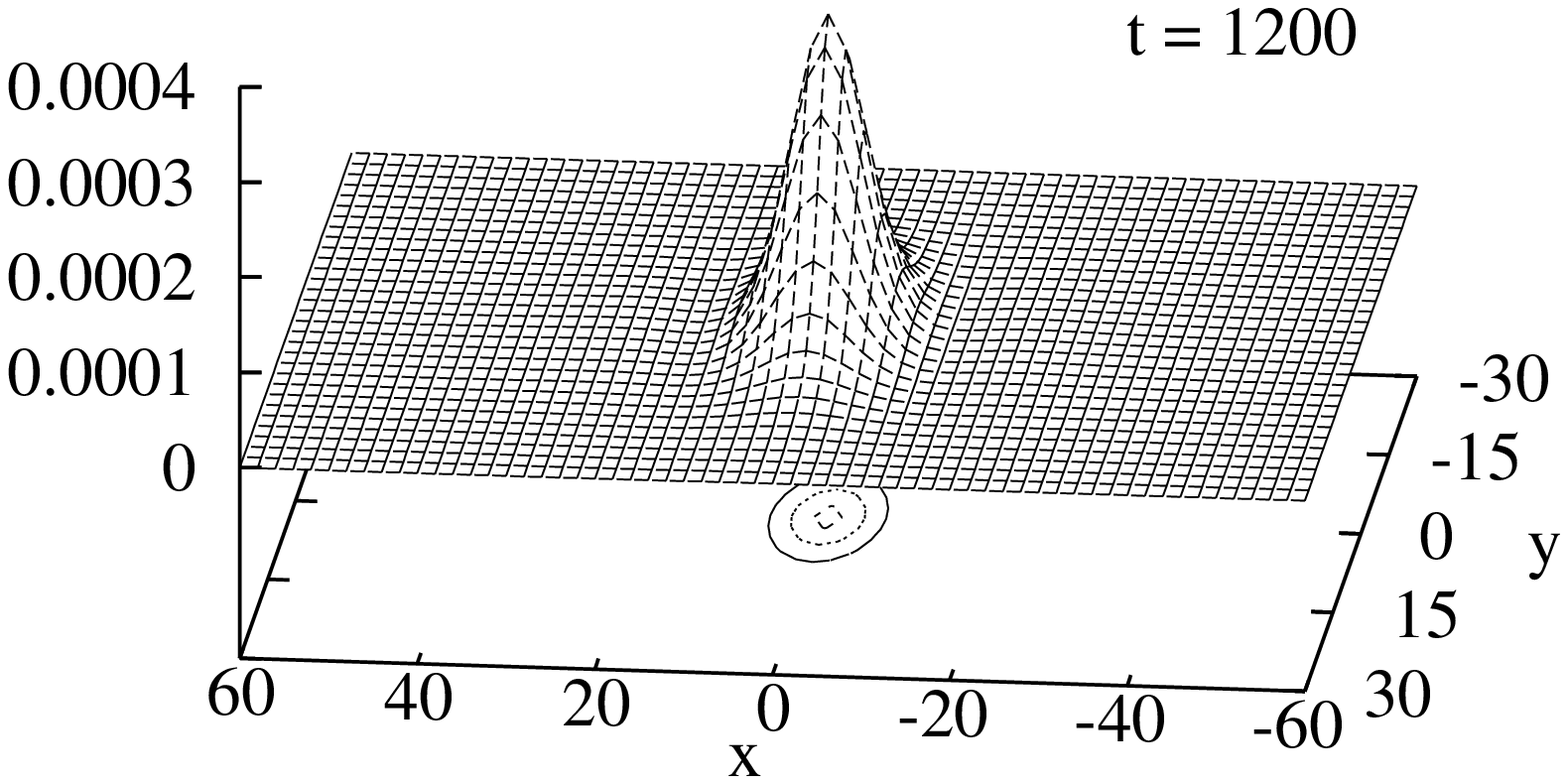,width=0.7\linewidth,clip=}
\end{tabular}
\caption{\textit{Boson/Boson pair}. 2D $z=0$ cuts of the Noether density $J^0$ at different times.  
As the stars come closer and merge, the maximum value of $J^0$ grows significantly followed by a quadrupolar
oscillation.}\label{2equal_Q}
\end{figure}

\begin{figure}
\centering
\begin{tabular}{c}
\epsfig{file=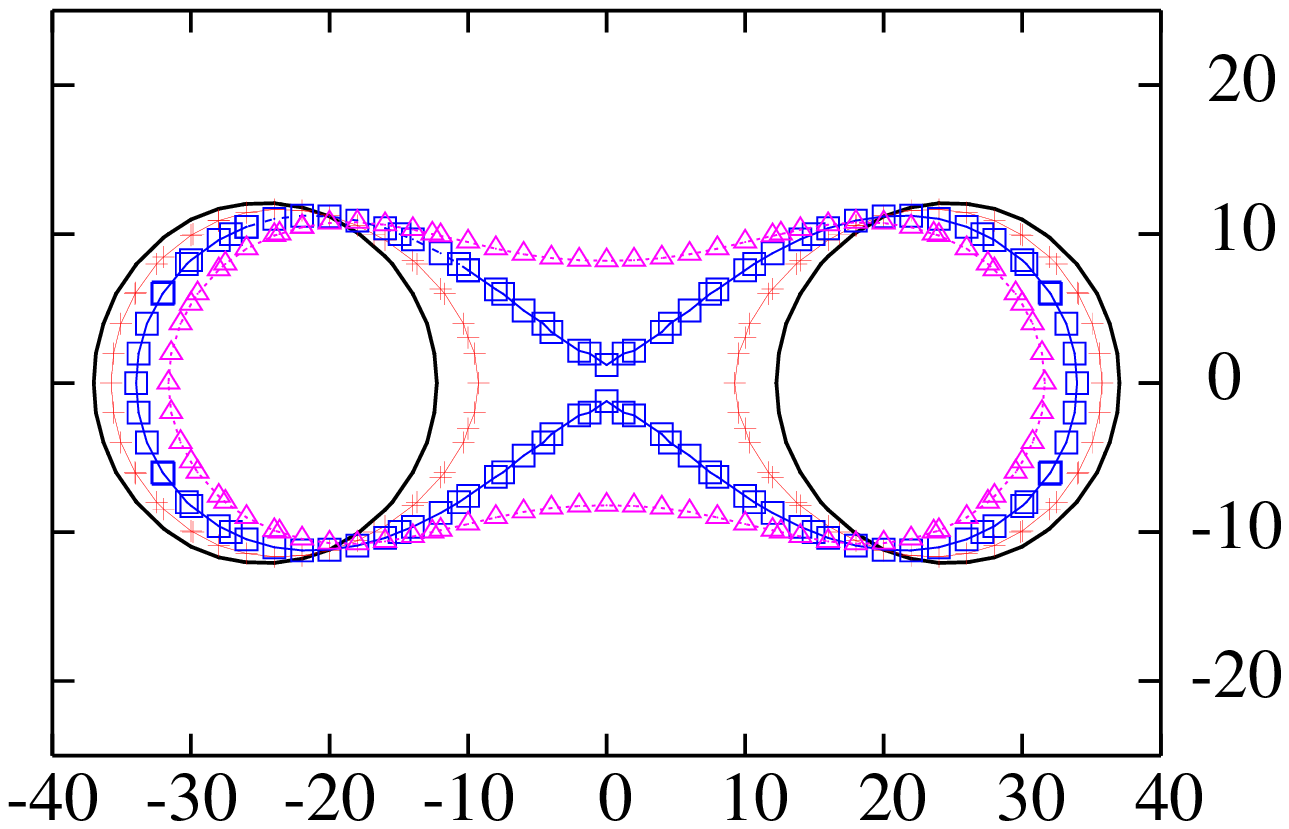,width=0.5\linewidth,clip=} \\
\epsfig{file=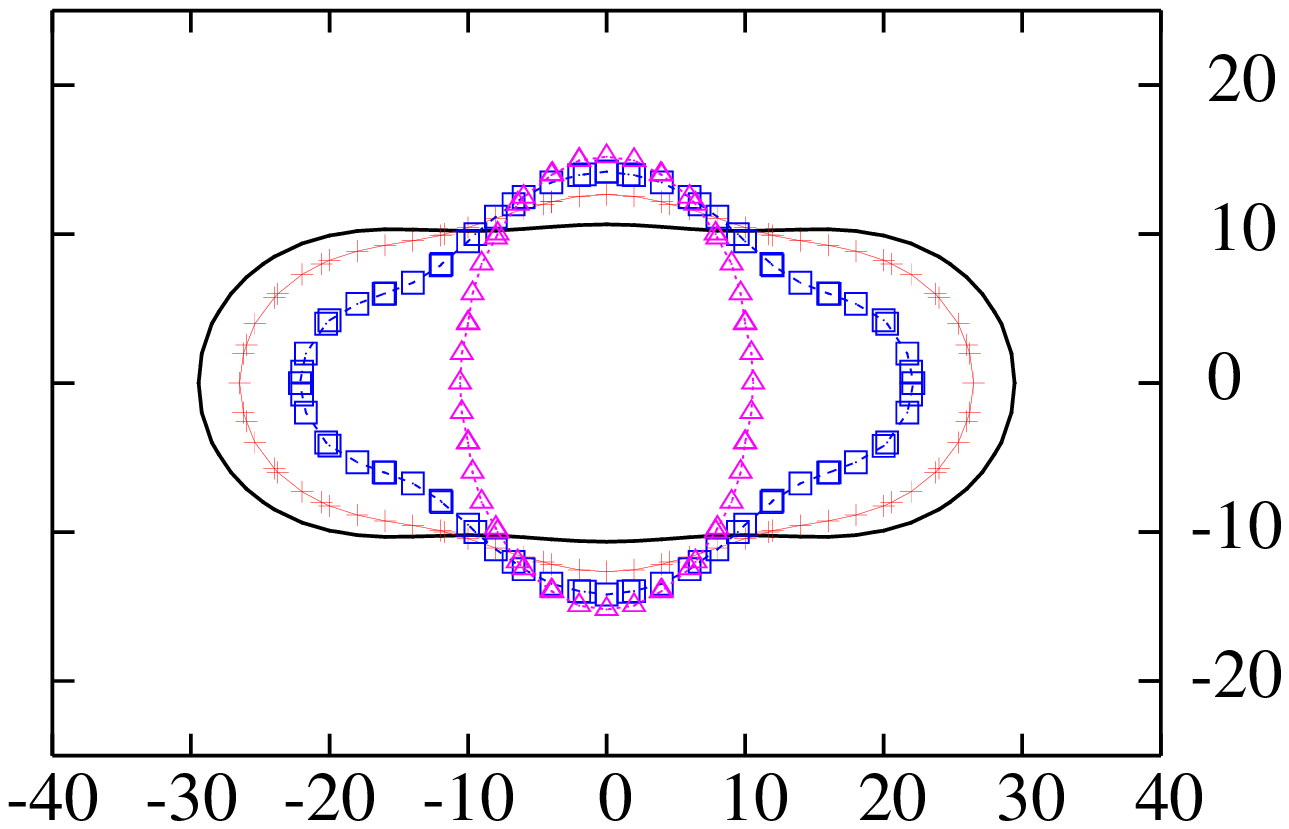,width=0.5\linewidth,clip=} \\
\epsfig{file=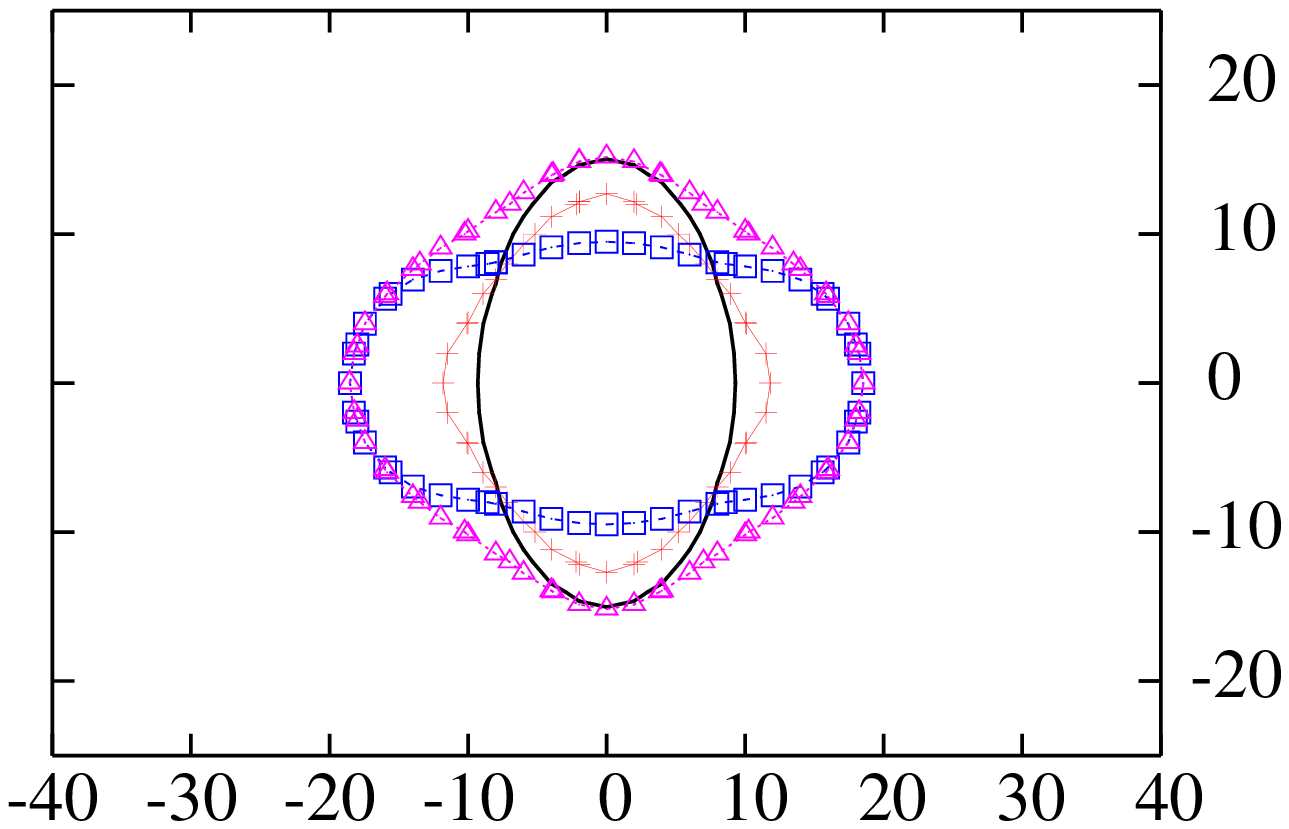,width=0.5\linewidth,clip=} \\
\epsfig{file=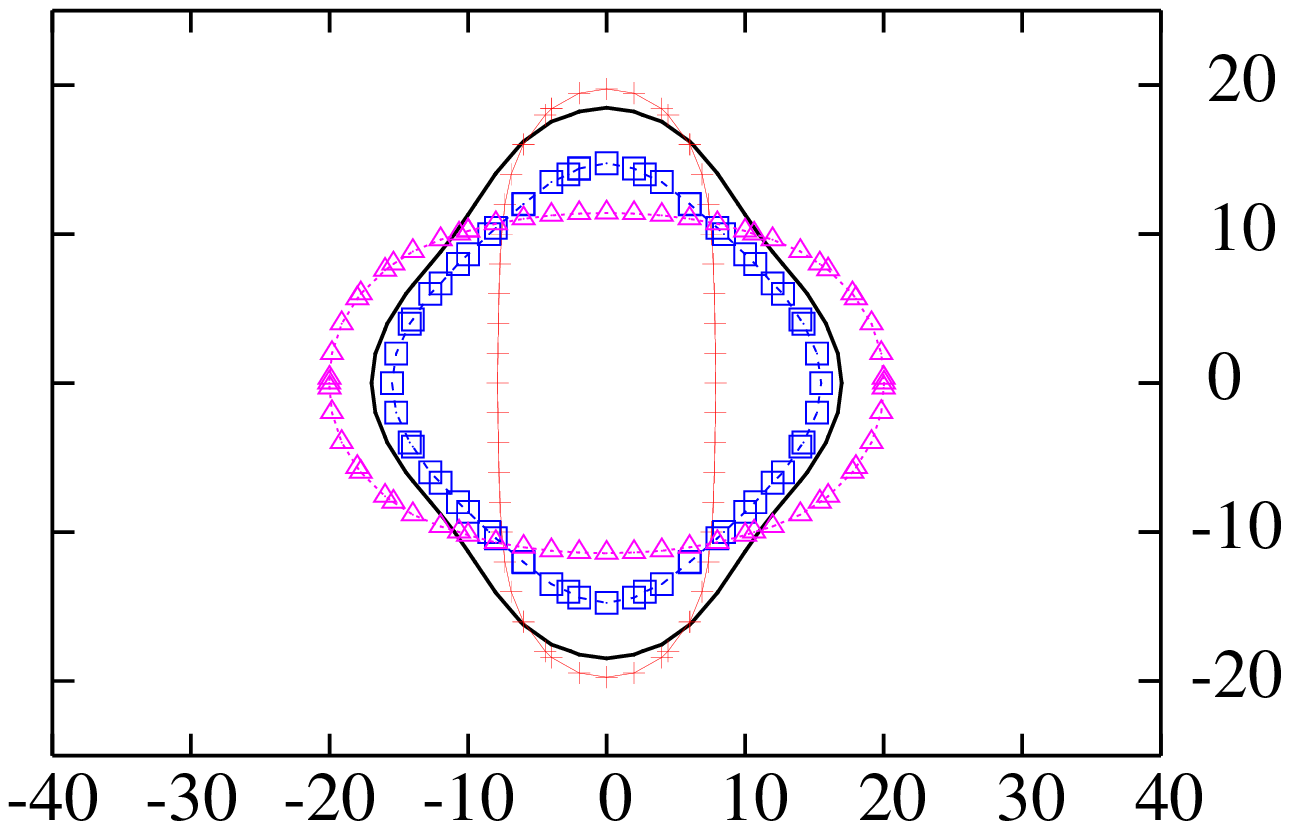,width=0.5\linewidth,clip=} \\
\epsfig{file=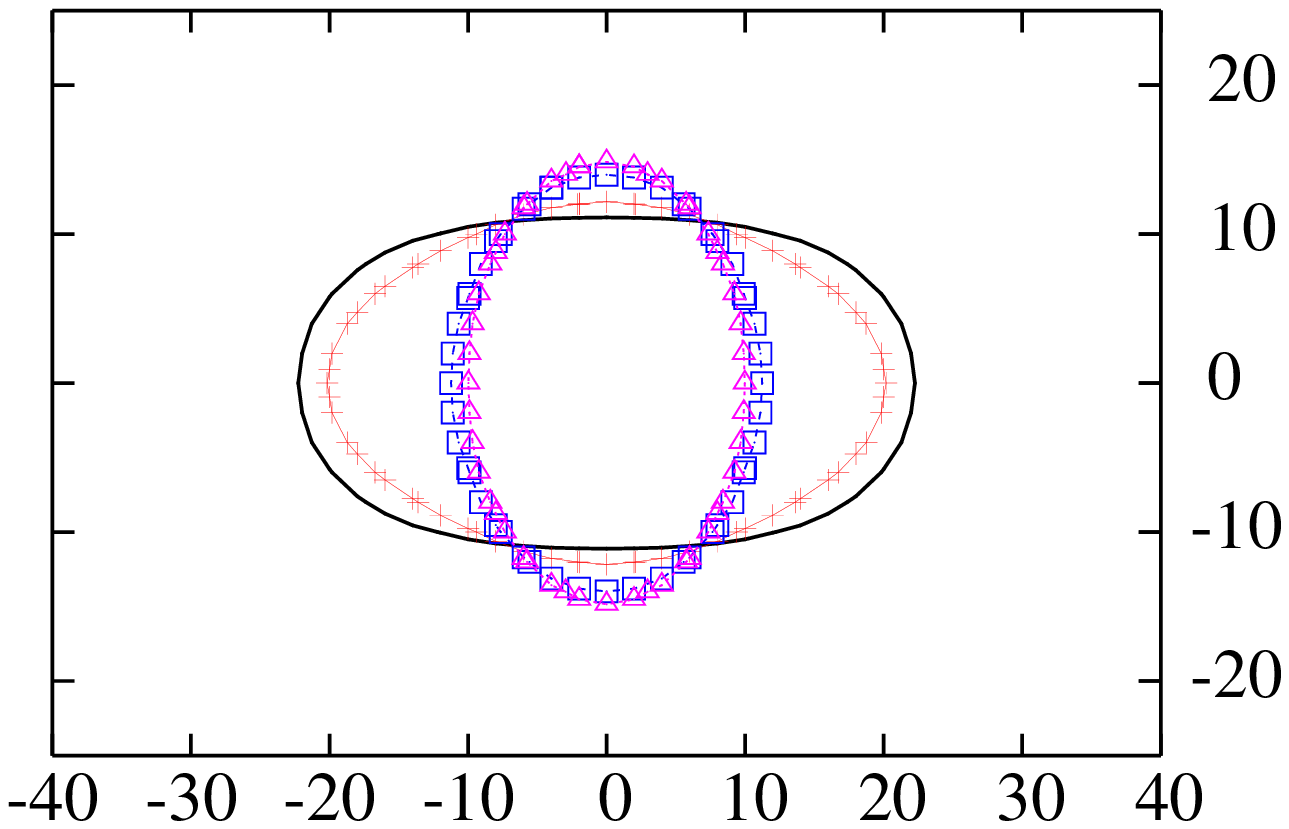,width=0.5\linewidth,clip=}
\end{tabular}
\caption{\textit{Boson/Boson pair}. Contours corresponding to the value $J^0 = 2 \times 10^{-5}$ 
at the $z=0$ plane. From top to bottom the contours are shown for times $\{ 0,160,240,300\}$,
$\{ 340,380,420,460\}$,$\{ 500,540,580,640\}$,
$\{ 680,760,840,900\}$ and $\{ 980,1060,1140,1200\}$ indicated by solid, solid-with-crosses, solid-with-squares
and solid-with-triangles respectively in each plot.   
As the stars get closer, the initially spherical contours deform until merging as
illustrated by a cusp on the top-most figure. Afterwards, $J^0$ exhibits essentially
quadrupolar-type oscillations.}\label{2equal_Q_contour}
\end{figure}

To illustrate the dynamics displayed by the solution, we present in 
Fig.~\ref{2equal_Q} and \ref{2equal_Q_contour} a sequence of plots illustrating the
behavior of the Noether density $J^0$. The extremes of this
function correspond, at least initially, to the centers of the stars and
as the evolution proceeds, the extremes give an indication as to the movement of the stars.  
Fig.~\ref{2equal_Q} presents two-dimensional slices (at $z=0$) of $J^0$. Notice that
the maximum value of $J^0$ is reached after the collision. This suggests that the boson stars have merged
and oscillations of the merged object are clearly distinguishable. 
Fig.~\ref{2equal_Q_contour} shows the nature of these oscillations in more detail by presenting
contour plots of $J^0=2 \times 10^{-5}$
at more frequent times.  

\begin{figure}[t]
\begin{center}
\epsfxsize=8cm
\epsfbox{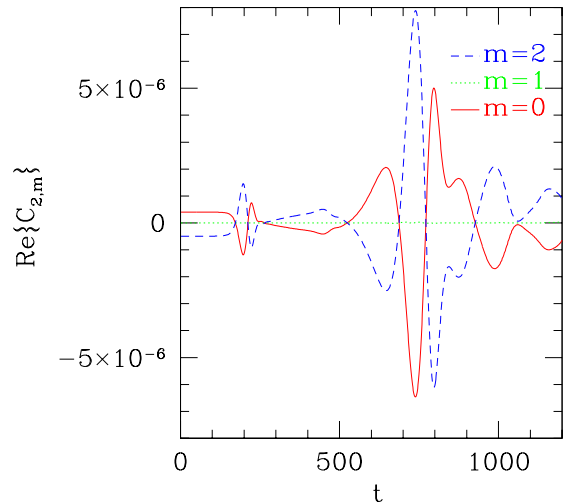}
\end{center}
\caption{\textit{Boson/Boson pair}. Coefficients corresponding to the $l=2$ modes of $r \Psi_4$ as a function of time, extracted at $r_{ext}=200$. After some initial transient due to spurious radiation in the initial data, 
the signal is clearly visible corresponding to the merger of the stars and later decaying to smaller
values. At late times, $t>800$, contamination with boundary effects obscures the extracted signals. 
}\label{psi4_modes}
\end{figure}

\begin{figure}[h]
\begin{center}
\epsfxsize=8cm
\epsfbox{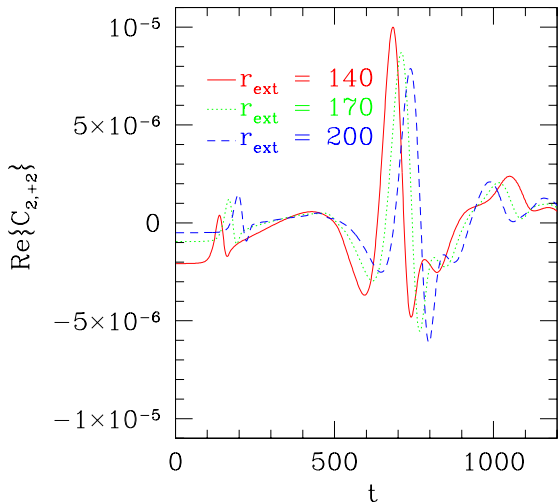}
\end{center}
\caption{\textit{Boson/Boson pair}. The coefficient $C_{2,+2}$ for different extraction radii. This 
figure illustrates both the outgoing waves due to the dynamics of the spacetime together
with incoming waves that have traveled to the boundary and bounced off it. Additionally
some remnant dependence on $r$ is visible indicating the extraction is still performed not
sufficiently far from the sources. Nevertheless the structure of the outgoing waves
is clearly visible.}\label{psi4_rext}
\end{figure}

The gravitational radiation produced in the collision, as encoded in $\Psi_4$,
is presented in Fig.\ref{psi4_modes} which illustrates
the (real) coefficients of its $l=2$ spin-weighted spherical harmonic modes.
Due to the symmetry of this particular problem and with the spherical
harmonics defined with respect to  the $z$ axis, there are the following  
simple relations between the non-trivial coefficients $C_{2,m}$,
\begin{eqnarray} 
   Re\{C_{2,2}\} &=& Re\{C_{2,-2}\} \\
   Re\{C_{2,2}\} &=& -\sqrt{3/2}~Re\{C_{2,0}\} ~~. \nonumber
\end{eqnarray}
Numerically we have found that $Re\{C_{2,2}\} = Re\{C_{2,-2}\}$ 
and $Re\{C_{2,2}\} = -1.22~Re\{C_{2,0}\}$,
which is in excellent agreement with the expected analytical relations  
The gravitational radiation emitted during the collision reaches the first
extraction surface (at $r_{ext}=140$) at $t=600$. It takes around $t=200-300$ to reach the boundary,
be reflected and pass again through the same extraction surface, as can be seen in 
Fig.~\ref{psi4_rext}. Notice that, despite having the surface extraction far from the sources,
they are still not well within the``wave-zone" since $r \Psi_4$ displays still a slight
dependence on $r_{ext}$. Nevertheless, the structure of the radiated wave is evident in
the plot. 

\begin{figure}[h]
\begin{center}
\epsfxsize=8cm
\epsfbox{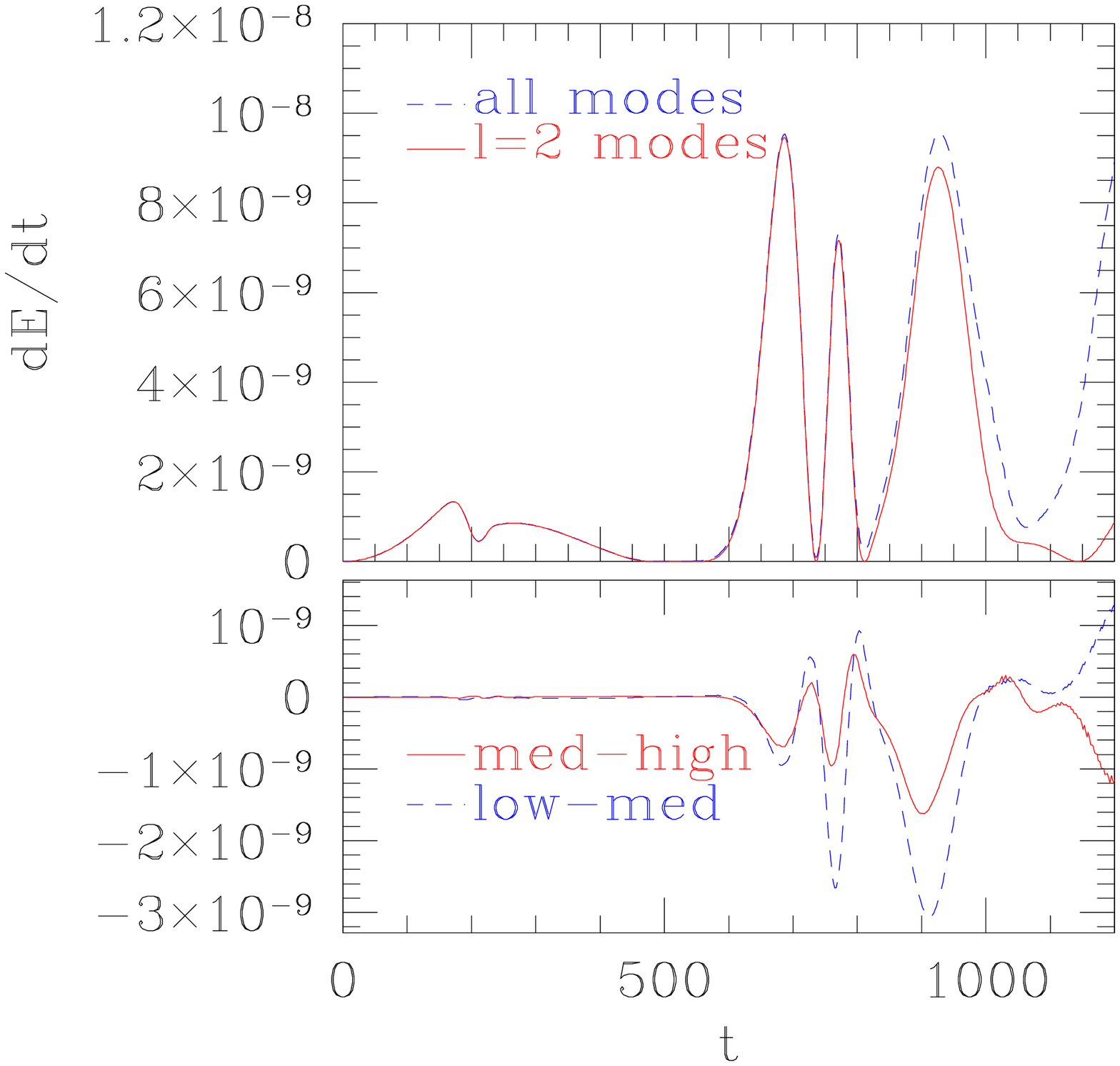}
\end{center}
\caption{ \textit{Boson/Boson pair}. Radiated energy as a function of time. The top plot indicates
the energy radiated in all modes (dashed line) together with that radiated solely in
$l=2$ modes. Clearly, until boundary effects begin to affect the results,
practically all energy is radiated in the $l=2$ modes. The bottom plot displays
the difference between the $l=2$ radiated energies at three different resolutions, indicating
convergent behavior until $t\simeq 950$.} \label{power}
\end{figure}

In Fig.~\ref{power} we have plotted the emitted energy (or power) $dE/dt$ (eqn. \ref{dEdt2}) as a function
of time for different resolutions. We find that practically all energy is radiated in the $l=2$ mode
which would simplify the parameterization of the obtained waveforms for their use in data analysis
searches~\cite{baumgarteetal}. 

These evolutions therefore suggest that the merger of the two boson stars produces an oscillating
single boson star. However, we note that the resulting star does not correspond to the boson
star with mass equal to the sum of the individual masses. One might suspect the collision dispersed
energy in the form of scalar and gravitational radiation, but that does not appear to be the case.
The ADM mass hardly decreases and the gravitational wave output is minimal. Instead, it would appear
that the resultant object has yet to settle into a stationary star.

\subsubsection{Collision of boson/boson in opposition of phase pair}\label{unequal_phase_merger}
The previous case considered, in its early stages, can be regarded as generic for equal mass objects
in a head-on configuration. The inner-structure details only become relevant when the objects
become sufficiently close to each other as indicated by the effacement theorem~\cite{damour,will}.

However, as the stars approach each other, the particular details of the star under
consideration can have strong consequences.
Let us consider the more general possibility of having a relative
phase difference between the single boson stars that we will use to construct the global solution
(\ref{ID_equalcase}), as for instance,
\begin{eqnarray}
  \phi^{(1)} &=& \phi_0(r)~e^{-i \omega t} \\
  \phi^{(2)} &=& \phi_0(r)~e^{-i (\omega t + \theta)}
\end{eqnarray}
with $\theta$ the relative phase. We will concentrate here on the extreme case, $\theta=\pi$ and
so the stars are in phase-opposition. Recall that while Einstein equations are insensitive to
this phase difference, the Klein-Gordon equations are. 
Notice that in our head-on configuration the surface $x=0$ is an anti-symmetry plane for the scalar field,
so it must remain zero in that plane. 
In the present case, the observed behavior can be regarded,
to a certain extent, as being influenced by a  ``repulsive" interaction of
two objects trapped in a gravitational potential well. This leads to the objects oscillating 
around the coordinate position $(x=\pm 15,y=0,z=0)$. Elucidating the final fate of this
system will require much longer evolutions (making sure there is no causal interaction
with the boundaries). Additional comments about this case in light
of result obtained in the other ones will be discussed in the conclusion section.

The gravitational radiation produced in this scenario is considerably weaker than the
previous case, as the involved objects never acquire significant velocities to induce a strong-time
dependent quadrupole. In fact, our extracted radiation is of the same order as the spurious signal produced by the
initial data. From these results coupled with the analysis presented in \ref{energy_considerations},
we conjecture that the maximum of the radiation happens
at $\theta=0$ while the minimum corresponds to $\theta=\pi$ being the radiation a smooth function 
of the phase difference.
However this must be substantiated by studying several representative cases.

\begin{figure}
\centering
\begin{tabular}{c}
\epsfig{file=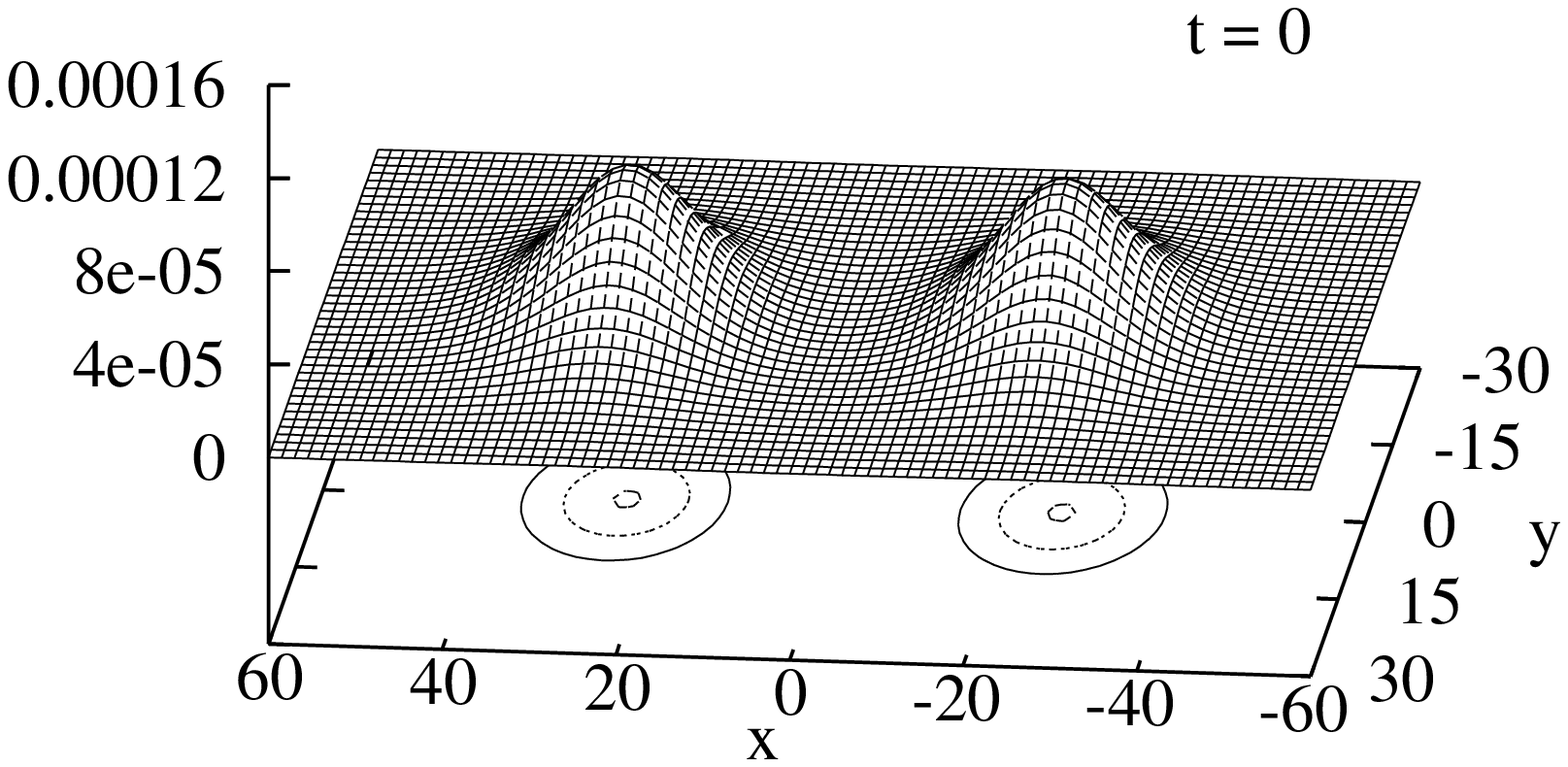,width=0.7\linewidth,clip=} \\
\epsfig{file=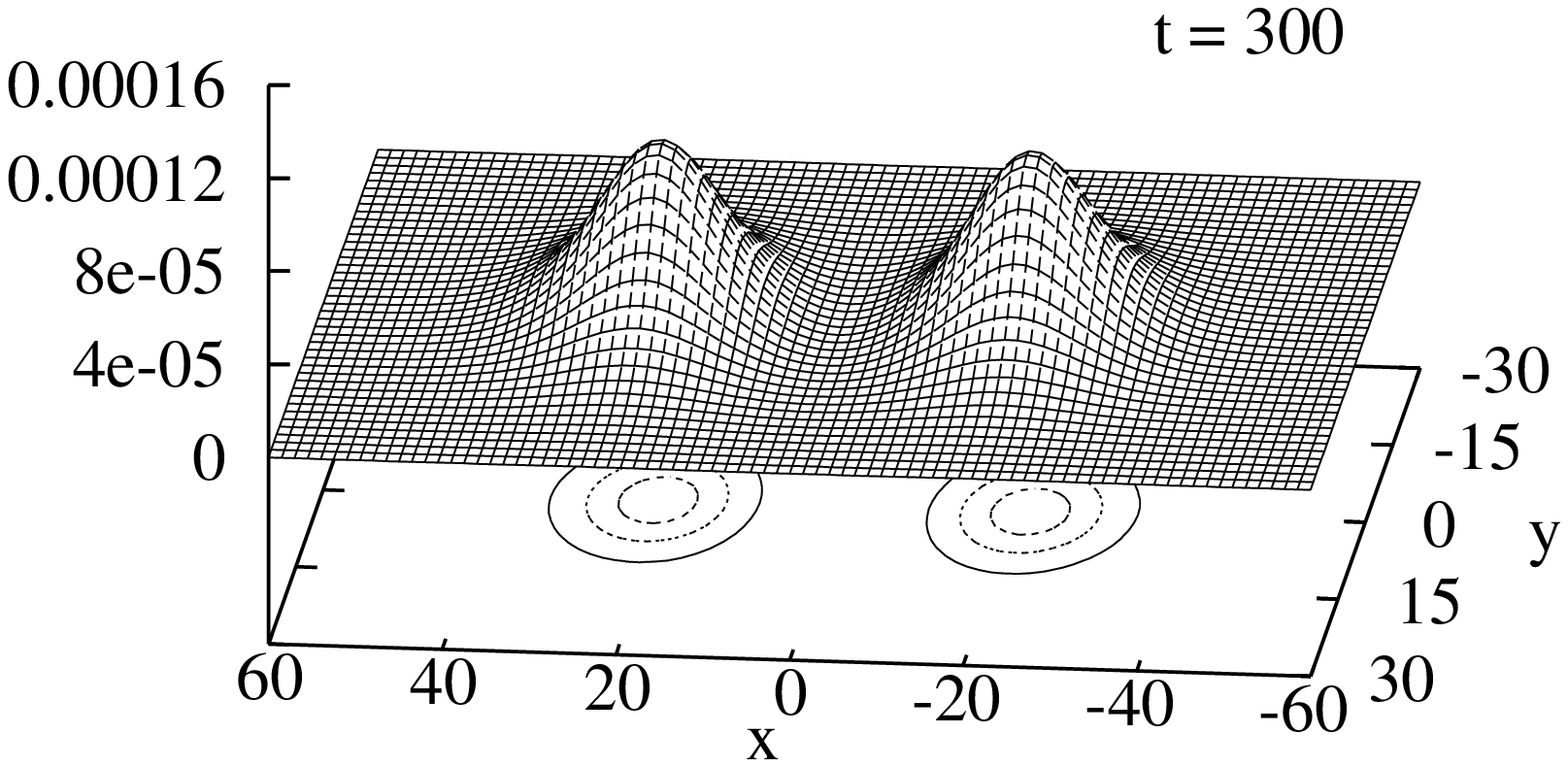,width=0.7\linewidth,clip=} \\
\epsfig{file=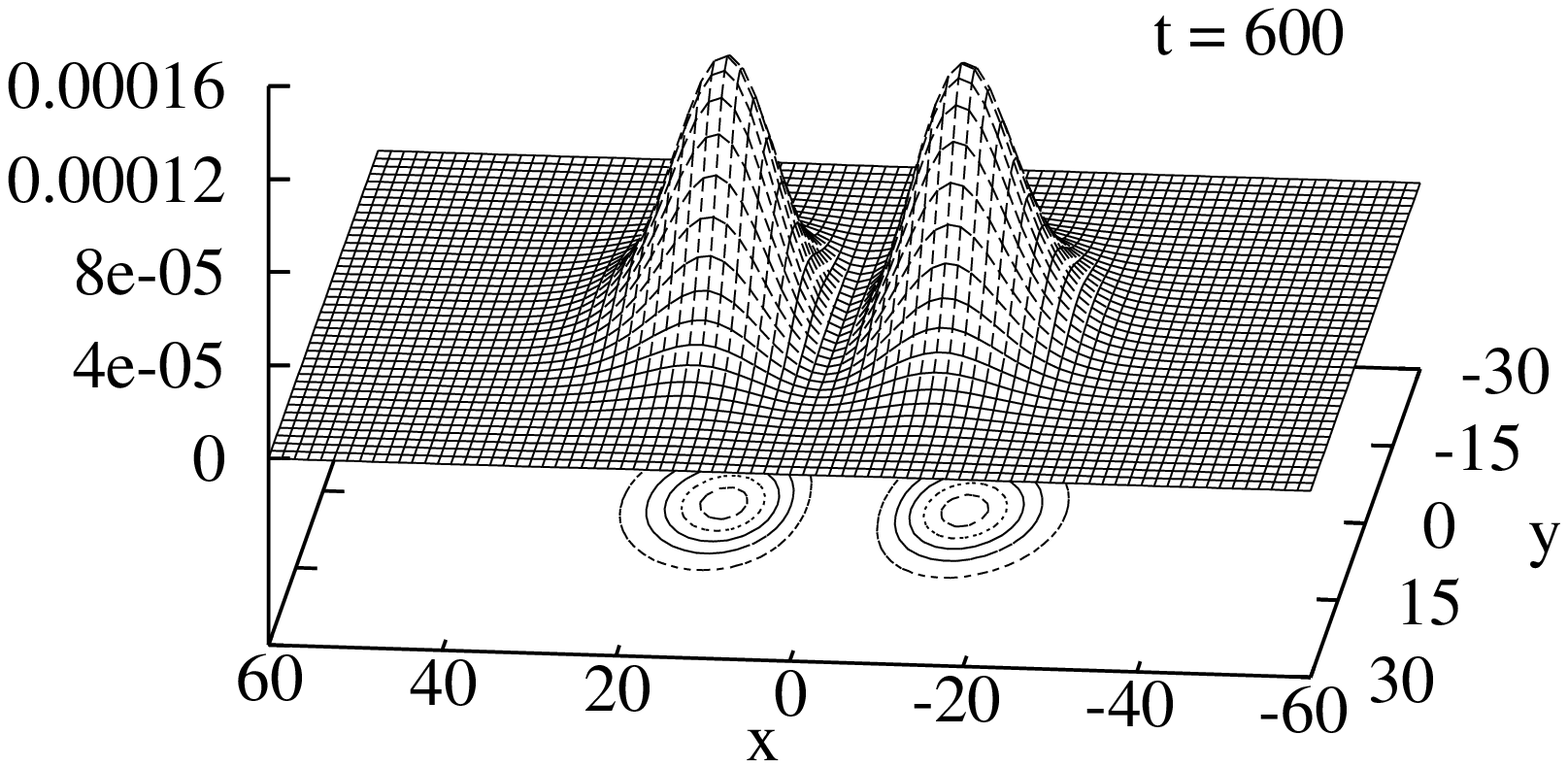,width=0.7\linewidth,clip=} \\
\epsfig{file=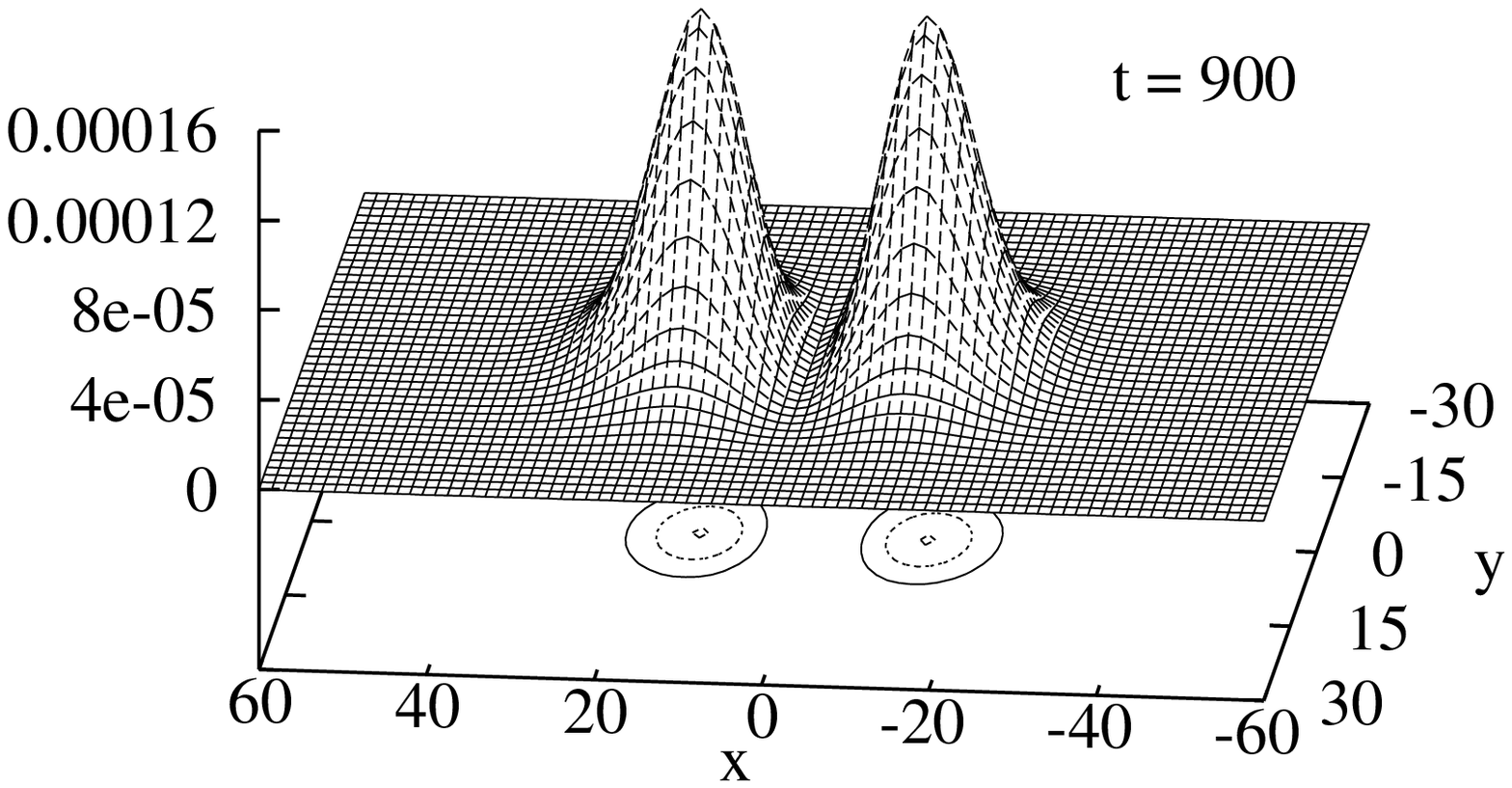,width=0.7\linewidth,clip=} \\
\epsfig{file=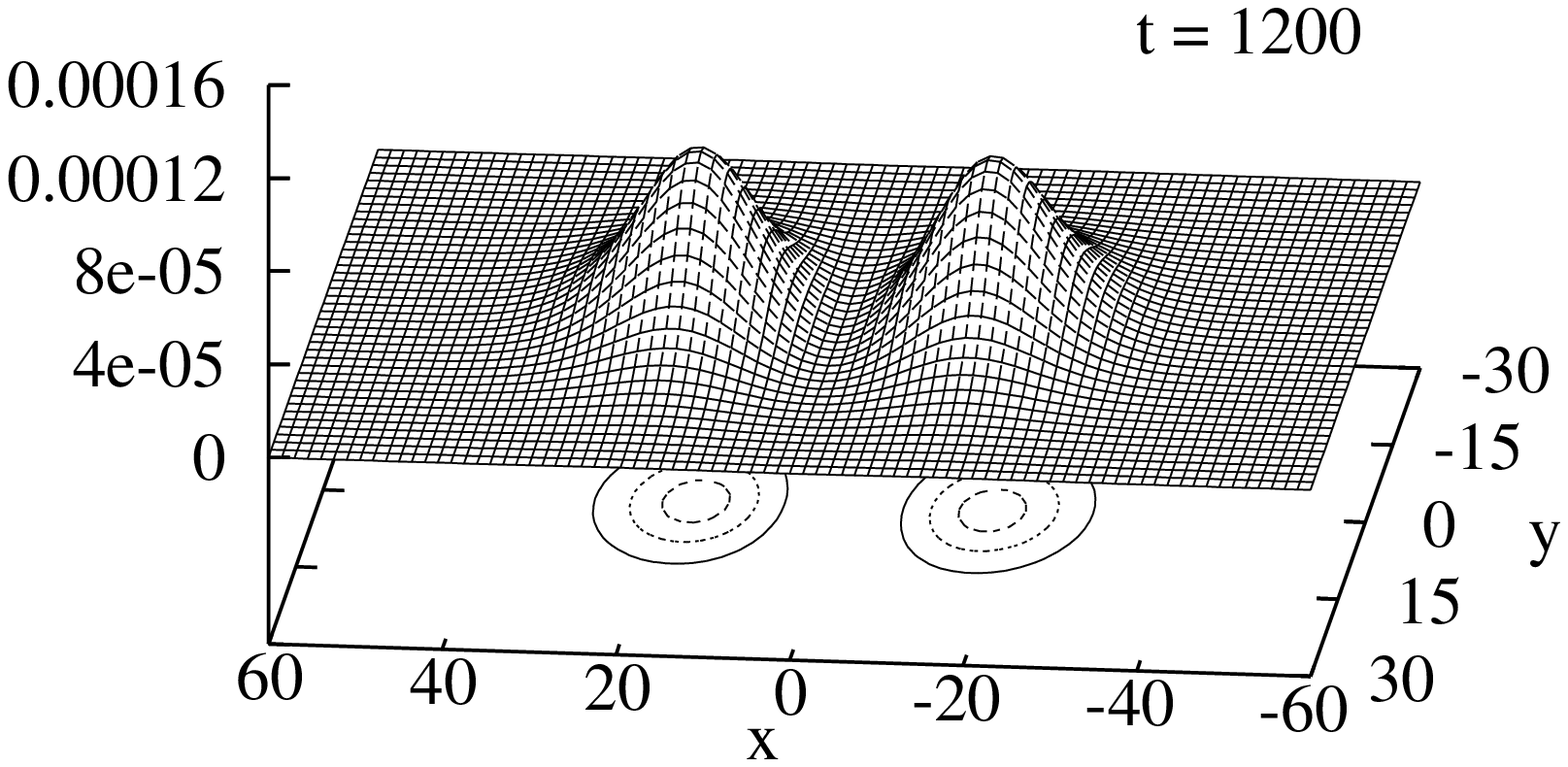,width=0.7\linewidth,clip=}
\end{tabular}
\caption{ \textit{Boson/Boson in opposition of phase pair}. 
 2D $z=0$ cuts of the Noether density $J^0$ at different times
Although the stars come closer, they never seem to merge.}\label{2phase_Q}
\end{figure}

\begin{figure}
\centering
\begin{tabular}{c}
\epsfig{file=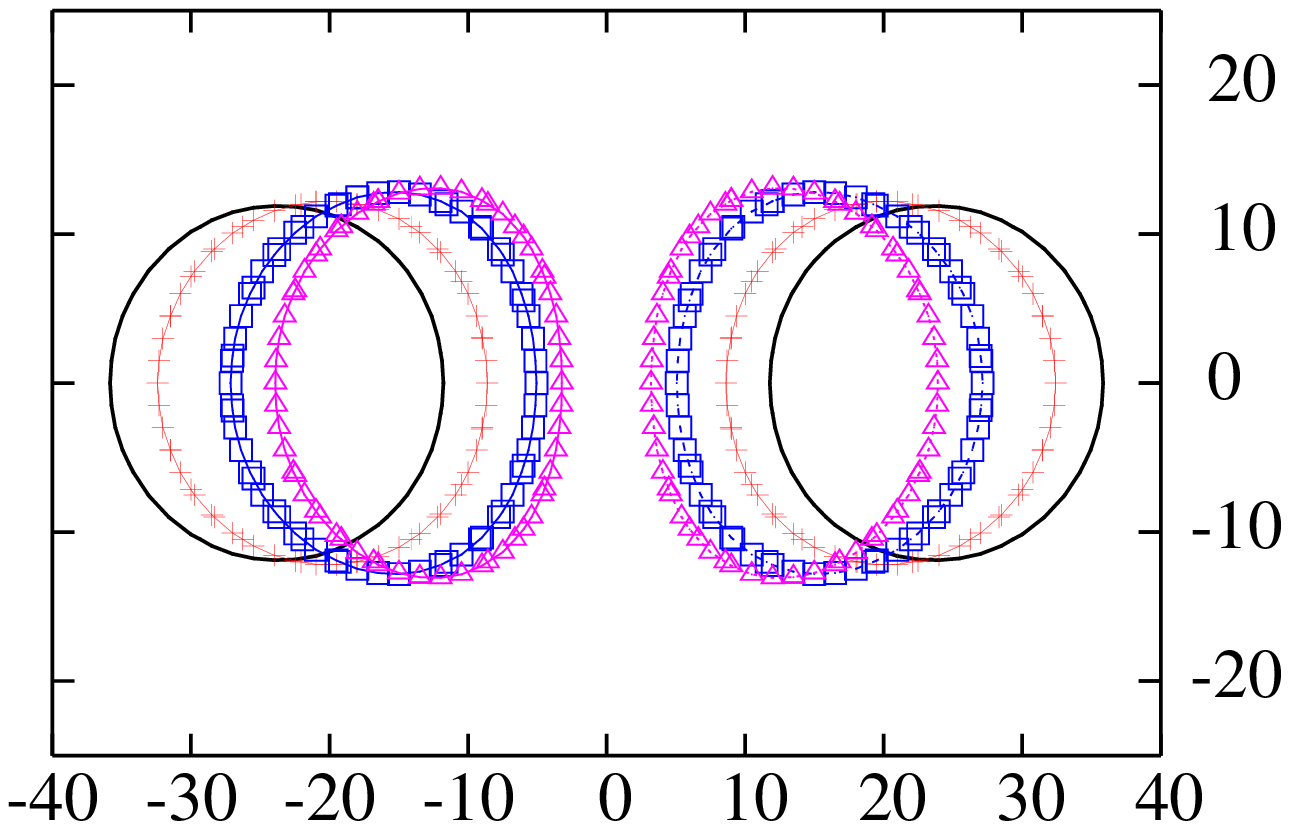,width=0.5\linewidth,clip=} \\
\epsfig{file=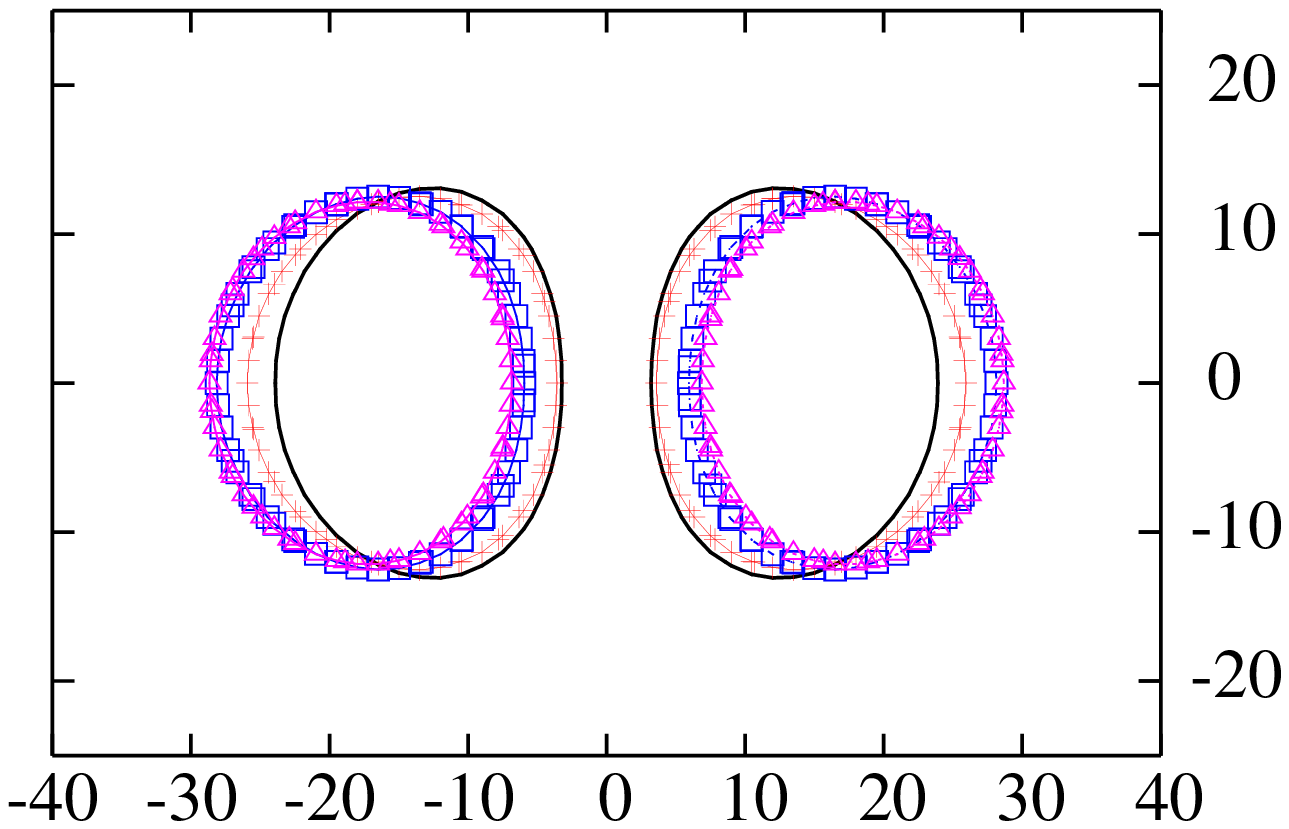,width=0.5\linewidth,clip=}
\end{tabular}
\caption{\textit{Boson/Boson in opposition of phase pair}. 
Contours with value $J^0 = 2 \times 10^{-5}$ on the $z=0$ place. From top to
bottom the contours are shown for times $\{ 180,360,540,720\}$ and $\{ 720,900,1080,1200\}$
indicated by solid, solid-with-crosses, solid-with-squares
and solid-with-triangles respectively in each plot.   
The stars initially get closer (top frame) but afterwards they move apart from each other (bottom frame),
this process continues on as the stars are trapped in a common gravitational well.
}\label{2phase_Q_contour}
\end{figure}

\subsubsection{Boson against antiboson}\label{boson_antiboson}

\begin{figure}[h]
\begin{center}
\epsfxsize=8cm
\epsfbox{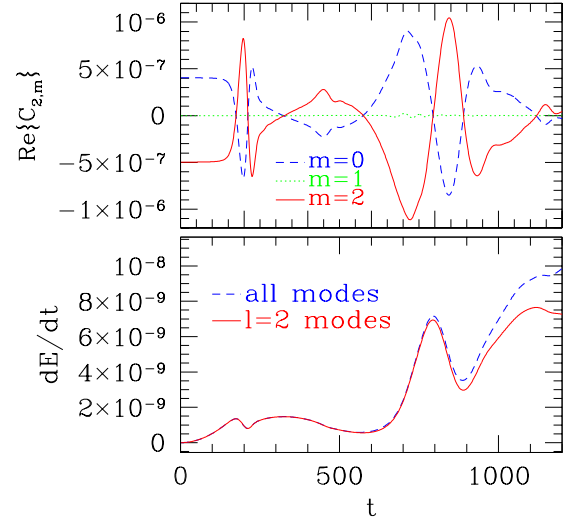} 
\caption{\textit{Boson/antiBoson pair}.  The $C_{2,m}$ coefficients of the $r \Psi_4$ decomposition (top frame) and the emitted radiation versus time (bottom frame).} 
\label{2anti_psi4}
\end{center}
\end{figure}

\begin{figure}
\centering
\begin{tabular}{c}
\epsfig{file=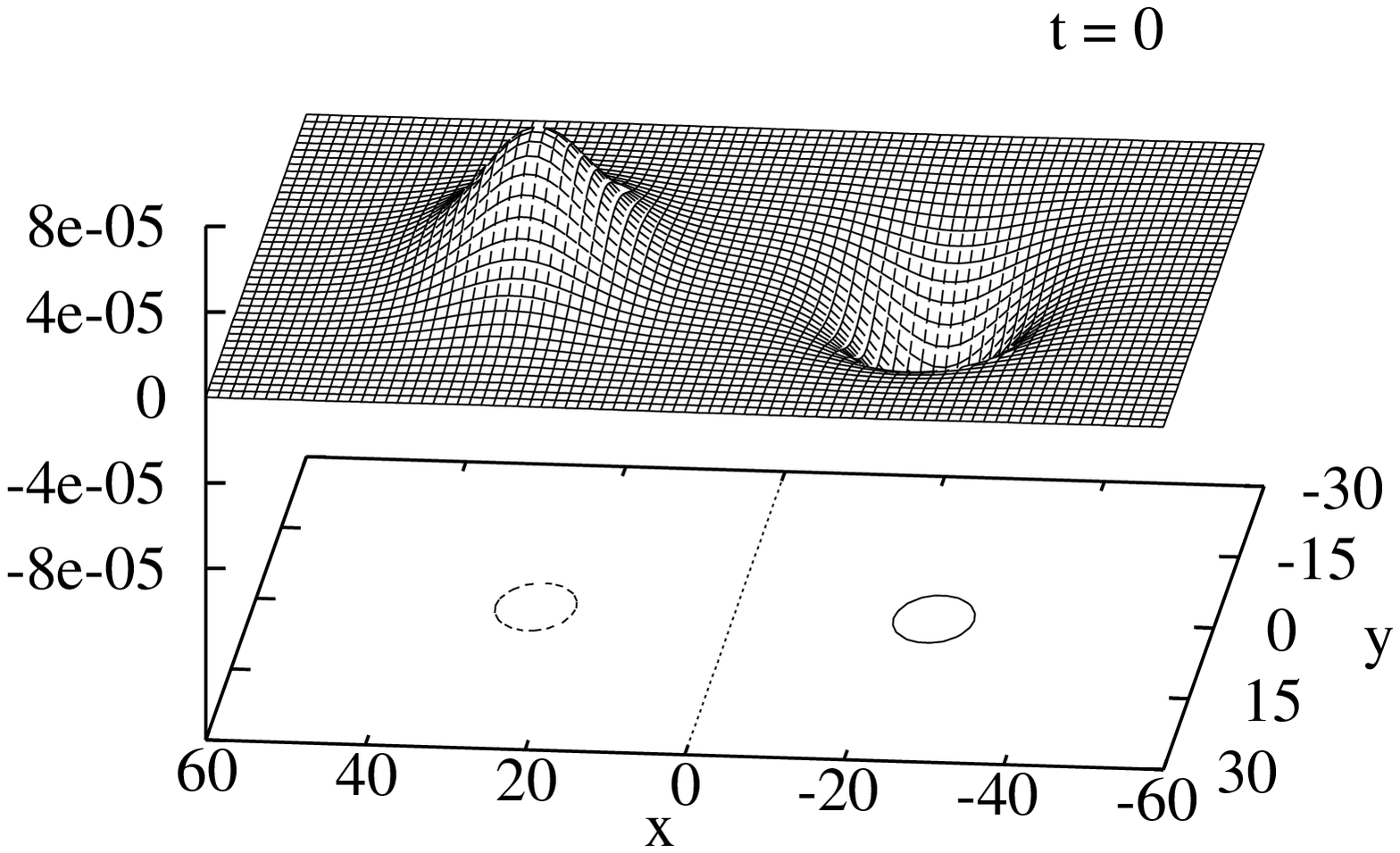,width=0.7\linewidth,clip=} \\
\epsfig{file=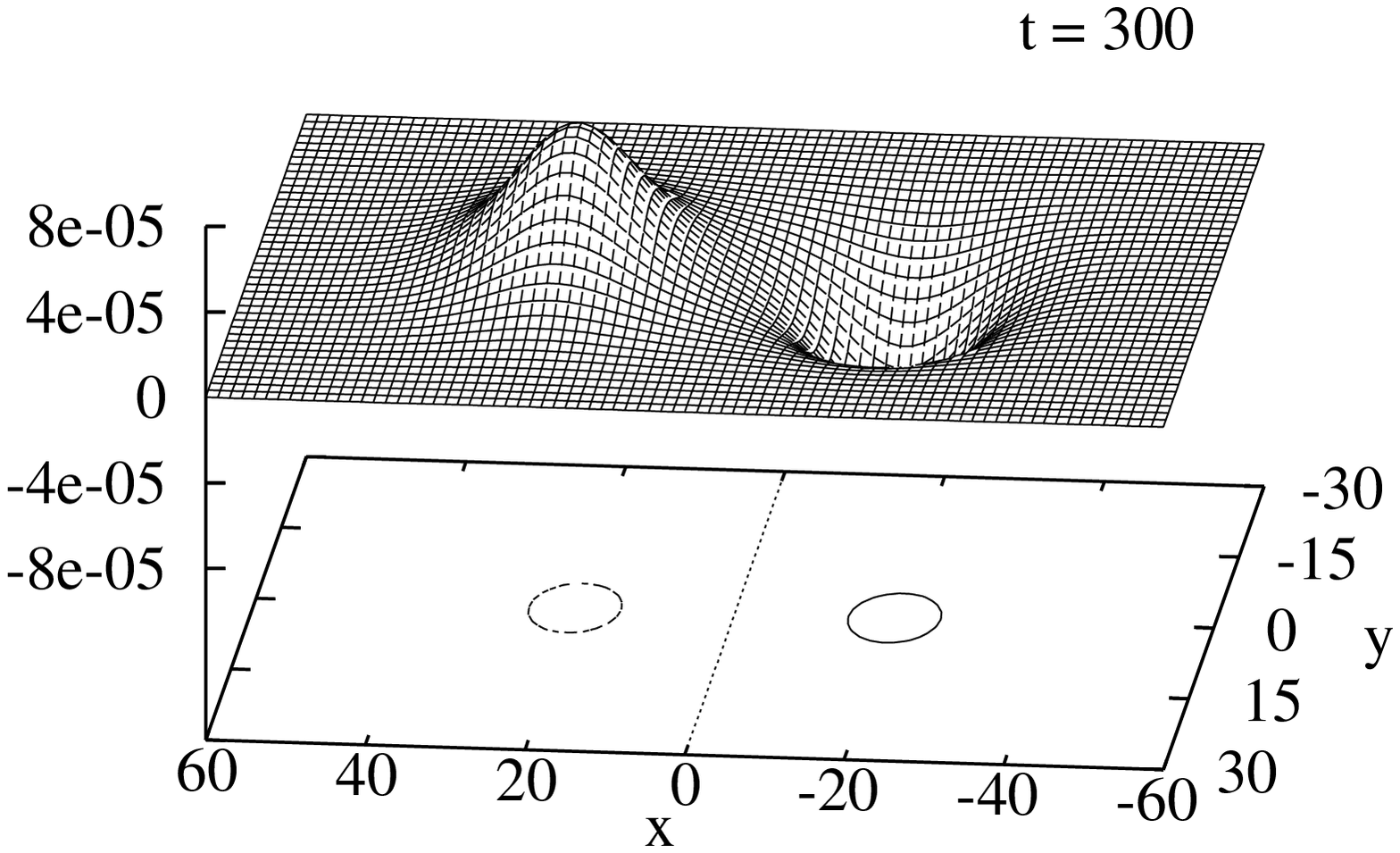,width=0.7\linewidth,clip=} \\
\epsfig{file=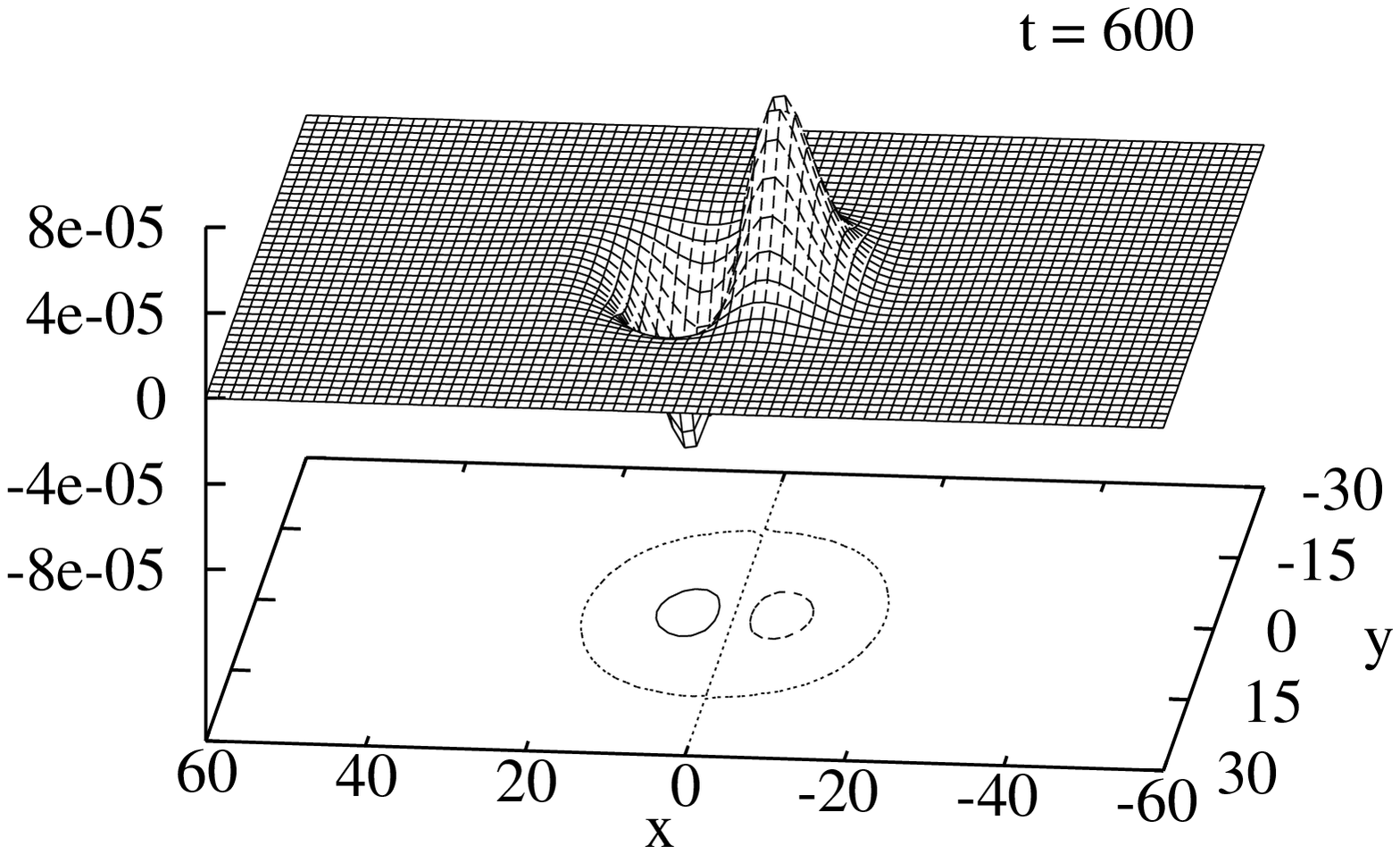,width=0.7\linewidth,clip=} \\
\epsfig{file=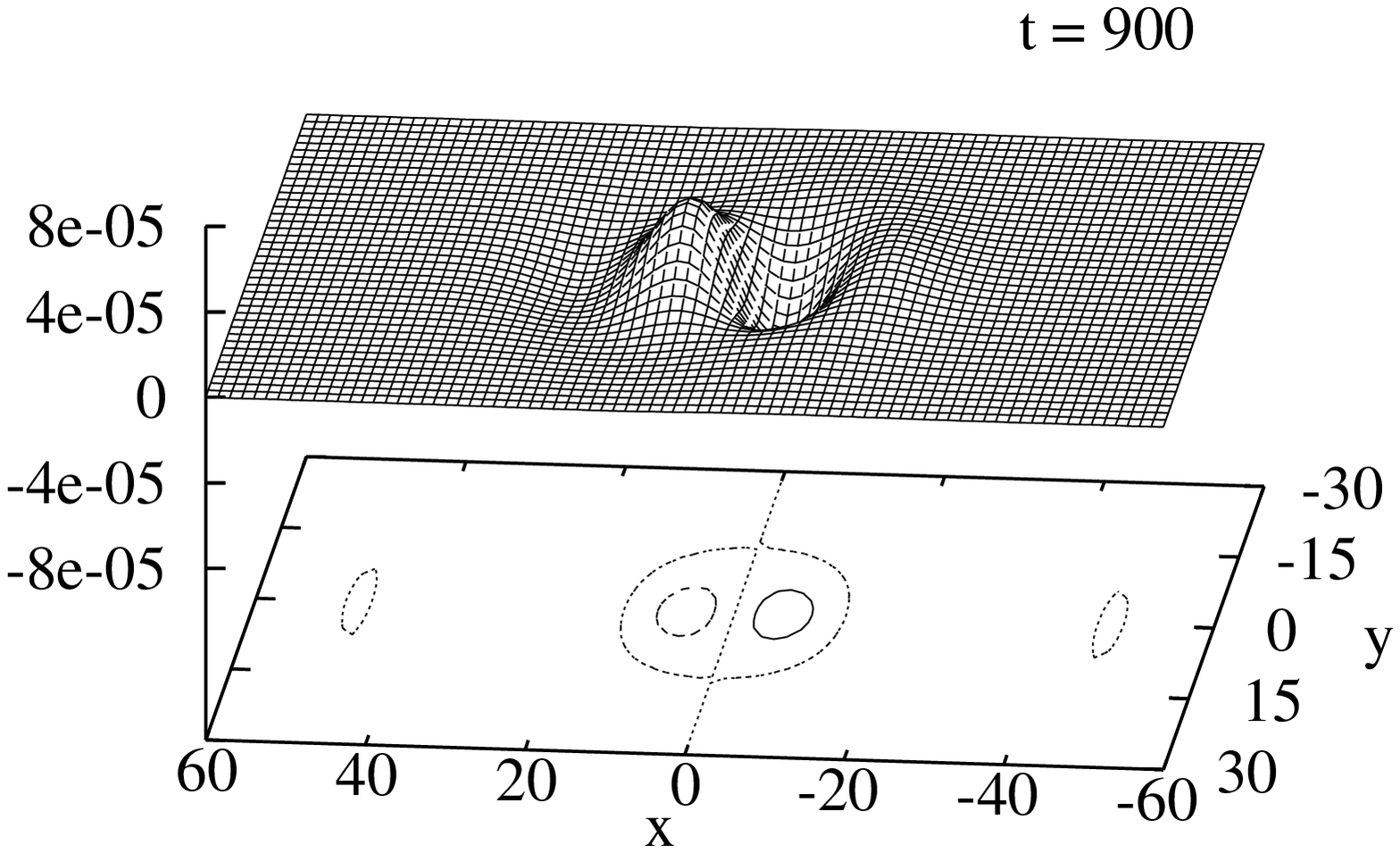,width=0.7\linewidth,clip=} \\
\epsfig{file=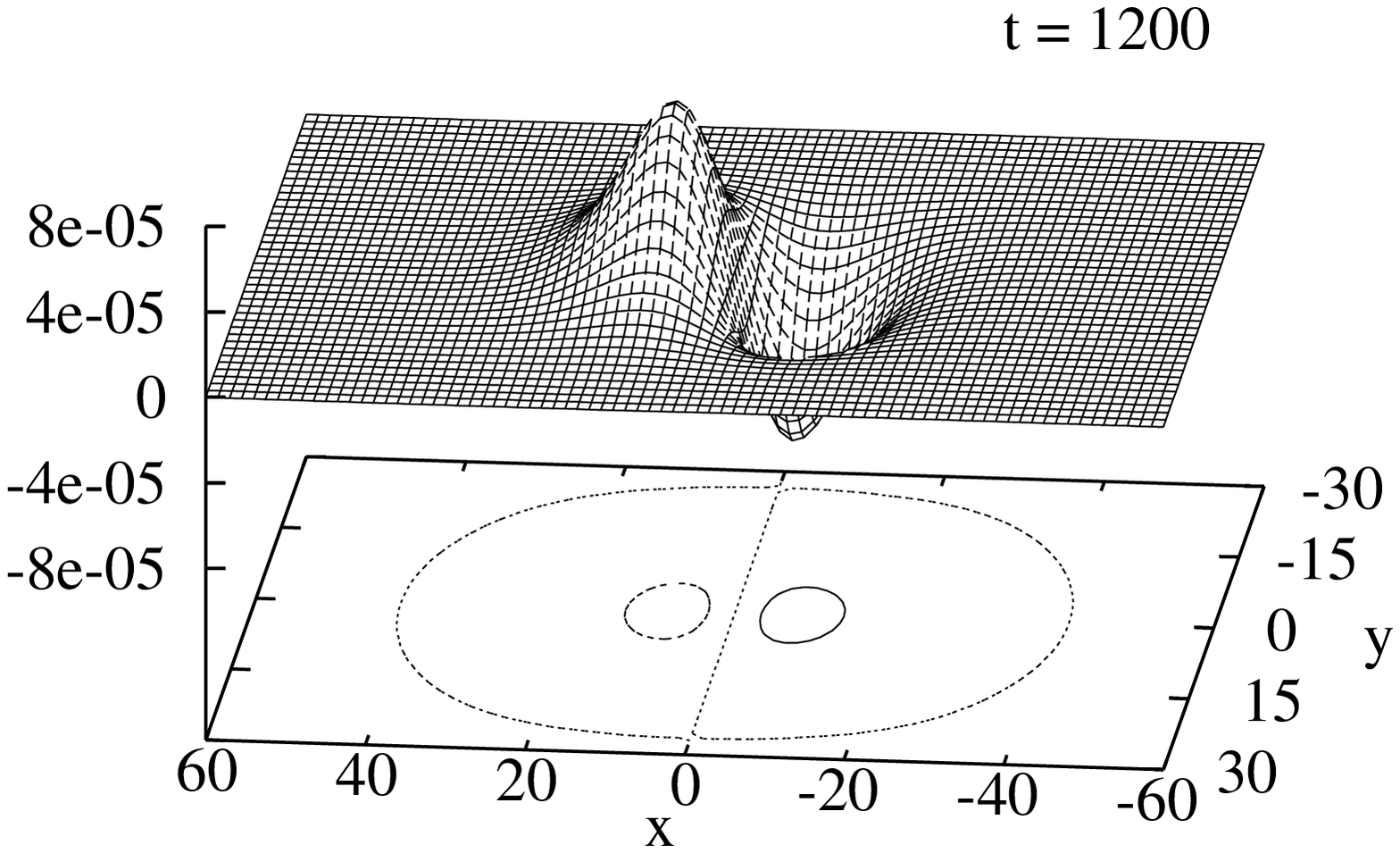,width=0.7\linewidth,clip=}
\end{tabular}
\caption{ \textit{Boson/antiBoson pair}. 
2D $z=0$ cuts of the Noether density $J^0$ at different times
Although the stars come closer and merge, the bosonic part is distinguishable from the 
antibosonic one by means of the different sign of the Noether density.}\label{2anti_Q}
\end{figure}

Another interesting case is the collision of a boson star with an otherwise identical star except with
the opposite charge density. Such a star is called an antiboson star and rotates in the opposite direction
as its partner in the complex plane.
Recall that the initial value problem solution was degenerate upon the reflection $\omega \rightarrow -\omega$.
Additionally, while this change leaves the geometry unchanged, the associated Noether change
is affected by an overall sign change. In this section, we study the dynamical behavior of a binary
system composed of a boson and an antiboson star. The initial conditions for such a scenario are simply
obtained with the following choice,
\begin{eqnarray}
  \phi^{(1)} &=& \phi_0(r)~e^{-i \omega t} \\
  \phi^{(2)} &=& \phi_0(r)~e^{+i \omega t} ~~.
\end{eqnarray}
We obtain the evolution of such a system, and as with the phase-opposition case, the early behavior agrees with 
that of the boson-boson case. As time progresses however, notable differences arise which are illustrated
in Fig.~\ref{2anti_Q}. The dynamical behavior when the stars are close is less strong than in the boson-boson case
but more so than in the boson-phase opposition case. As a result, there is non-trivial generation of gravitational
waves though their time-scale is longer than and their strength smaller than the boson-boson case. 
The total radiated energy however, is similar as illustrated in Fig.~\ref{2anti_psi4}.

\begin{figure}
\centering
\begin{tabular}{c}
\epsfig{file=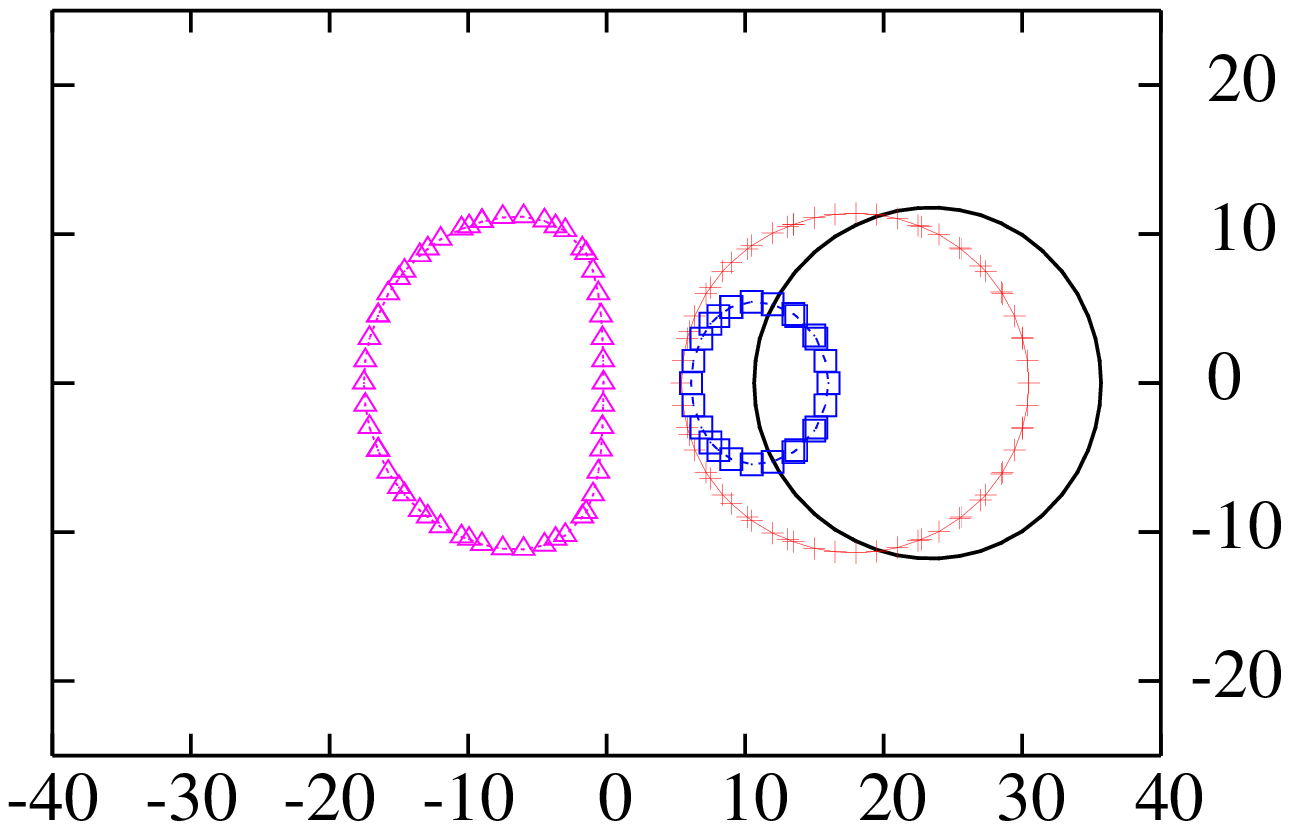,width=0.5\linewidth,clip=} \\
\epsfig{file=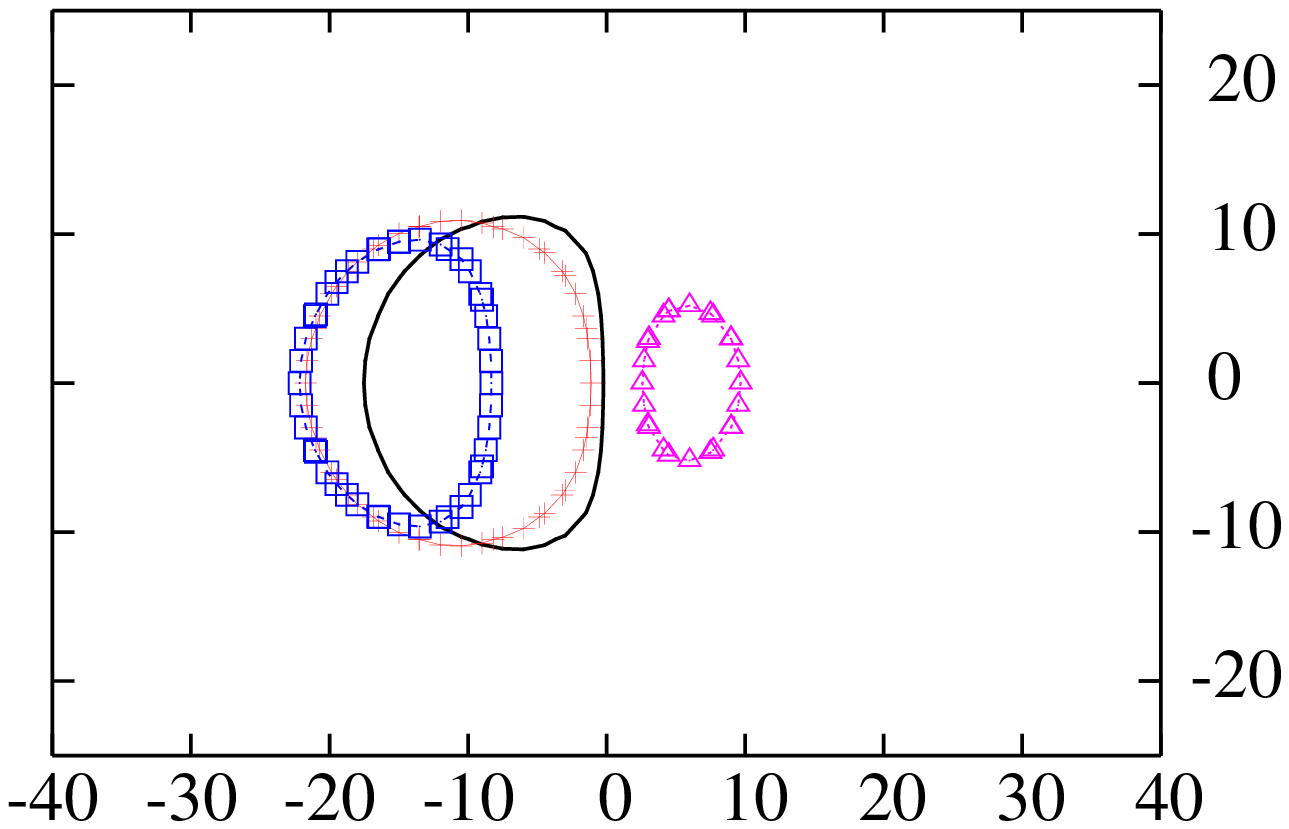,width=0.5\linewidth,clip=}
\end{tabular}
\caption{ \textit{Boson/antiBoson pair}. 
Contours with value $J^0 = 2 \times 10^{-5}$ on the $z=0$ place. From top to
bottom contours are shown for times $\{ 180,360,540,600,660\}$ and $\{ 660,780,840,900\}$
indicated by solid, solid-with-crosses, solid-with-squares
and solid-with-triangles respectively in each plot.   
Only the right star has such a contour initially as the left has
a negative initial value for $J^0$. As the stars come closer to each other, the
``tunneling'' behavior is illustrated by the positive value appearing on the left side (top frame).
As time progresses, this process essentially reiterates itself from the left side to the
right side (bottom frame). The dynamics that follows repeats this side to side `motion' in the
Noether density with the stars trapped in a common gravitational well.}
\label{2anti_Q_contour}
\end{figure}


\section{Conclusions}
\label{conclusions}
We have studied the behavior of boson stars both
in isolation and head-on collisions with a three-dimensional
implementation of the Einstein equations.
Studies of the  isolated star in Section \ref{single_boson} 
provided a non-trivial test of the implementation
and showed it capable of accurately extracting delicate features
of the solution such as the oscillation frequency. Additionally by considering
a coordinate condition not adapted to the static spacetime under study,
the solution obtained further illustrated the ability of the harmonic coordinates
to deal with compact objects in a smooth manner.

We then considered head-on collisions of different configurations
of boson stars. Our studies revealed the interesting and varied phenomenology
interacting boson stars can give rise to. Additionally, these scenarios provided
an excellent test of the implementation under strongly dynamical conditions. This 
implementation is being employed to deal with black hole spacetimes, and the
results will be communicated elsewhere. 

\begin{figure}[h]
\begin{center}
\epsfxsize=8cm
\epsfbox{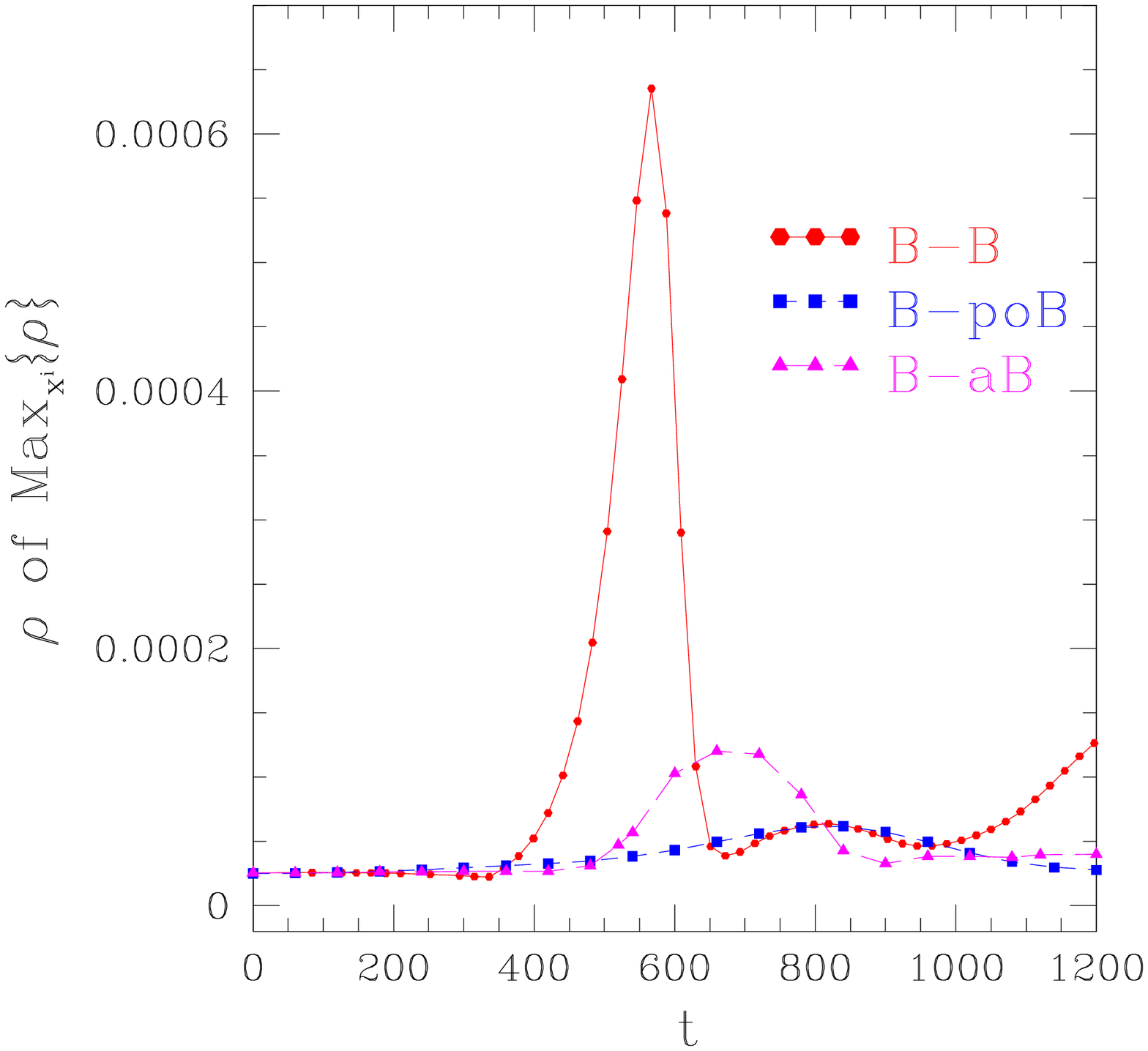}
\end{center}
\caption{The maximum of the energy density $\rho$ is plotted as a function of time for the
three different cases. }\label{maximums}
\end{figure}

\begin{figure}[h]
\begin{center}
\epsfxsize=8cm
\epsfbox{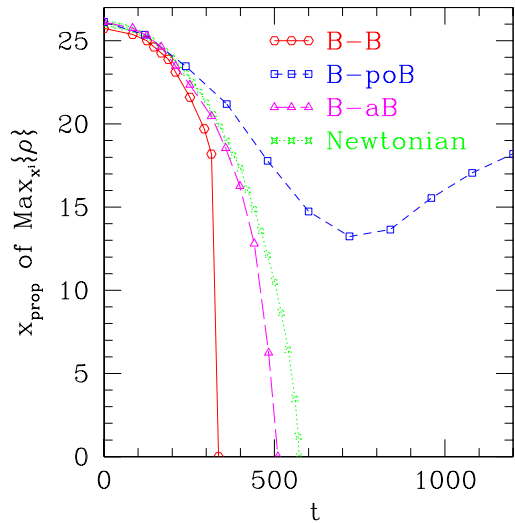}
\end{center}
\caption{The proper distance from the center of the boson star to the center of the 
computational domain as a function of time for the
different cases and the Newtonian approximation. The position of the boson is identified
with the maximum of the energy density.}\label{centers}
\end{figure}

In addition to the notable differences obtained,
it is interesting to pay a closer look
at the position of the boson stars as a function of time. 
The positions are determined by the location where the energy density
reaches a maximum in a given neighborhood. Fig.~\ref{maximums}
shows the maximums of the energy density $\rho$ while Fig.~\ref{centers}
shows the proper distance between the origin and
these maximums (i.e., that can be identified with the center of the boson stars)
located on the $+x$ direction. Due to the symmetry of the problem this is enough 
to draw conclusion on both stars.  Included in the figure is the
position a star would have as dictated by a simple Newtonian behavior of
two point-particles with the mass of the boson star.

A priori, since the initial separation of the stars is of the same order of their radii,
non-Newtonian effects would be expected to become relevant early on. 
This is clearly visible for the boson-boson and boson-boson in phase 
opposition cases, however, the Newtonian result is a reasonably good 
approximation for the boson-antiboson collision.  This results from the 
significantly different interaction between their scalar fields in these cases, 
while in the boson-boson case a merger takes place at relatively early times, in the
boson-boson in phase opposition the stars approach and essentially go through each-other.

A further remark about the particular cases that we have evolved is that the difference 
between the coordinate and the proper distances, during most part of the evolution (the early part), 
is given in all cases by a slowly varying function of time, compared to the time scale of the solution.

Additionally, these plots also explain, qualitatively, why the radiated energy is stronger for the 
boson-boson case since it is the one where the source's quadrupole has the most dynamic behavior.
It would be tempting to regard this as further evidence that in rather simple scenarios,
the coordinate behavior provides a reasonable description of the objects dynamics as seen
in the context of equal-mass binary black hole cases 
\cite{buonano,Campanelli:2006gf,Baker:2006yw,Bruegmann:2006at}. However one must be 
cautious in the present case as we are considering a particular model to define the 
scalar fields through a single complex field.

Also, interesting differences are observed when comparing our results with those 
of the study of Q-balls~\cite{Qball}. Here, the inclusion of gravitational effects
modify the dynamics strongly. Besides not needing to boost the stars for
a collision to take place, the gravitational attractions of the stars in our case 
is evidenced by the ``trapping''  of the stars instead of ``bouncing apart'' 
(for the boson-boson in phase opposition) or going through each other 
(for the ``boson-antiboson'') case as in~\cite{Qball}.  

Finally, the results presented here illustrate the rich behavior displayed
by binary boson star dynamics even in the rather simple scenario of a head-on
collision. In particular, the dependence on possible phase differences revealed
these phases strongly influence the outcome of the problem. In 
appendix \ref{energy_considerations} we present a simple treatment that sheds light
in the influence different phases or frequency-reflection in one of the stars' description have in
the dynamics of the problem.
In particular, this treatment suggests that  energy
arguments explain why the boson-boson case gives rise to a merge while in the boson-opposition 
in phase boson case they seem to repel each other.

Having examined the landscape of head-on boson star collisions behavior,
near future studies will concentrate on the more complex situations of
orbiting cases as well as boson-star black hole binaries.

\acknowledgments
We would like to thank M.~Anderson, C.~Bona, D.~Garfinkle, E.~Hirschmann, D.~Nielsen, 
F.~Pretorius, J.~Pullin and R.~Wald for helpful discussions.
This work was supported in part by NSF grants PHY-0244699, PHY-0244335,
PHY0326311 and PHY0554793  to Louisiana State University and PHY-0325224 to Long Island University.
The simulations described here were performed on
local clusters in the Dept. of Physics \& Astronomy at LSU and the
Dept. of Physics at LIU as well as on Teragrid resources provided by
SDSC under allocation award PHY-040027.
L.L is grateful to the Alfred P Sloan Foundation and
Research Corporation for financial support.

\appendix
\section{Initial Data for an Isolated Boson Star}
\label{initial_data}
The initial data for the boson star configuration is computed first
in spherical symmetry with a one-dimensional code. The resulting
solution is then extended to three dimensions. The one-dimensional solution
is obtained in the following way. First, we
adopt a specific ansatz for the scalar field,
\begin{equation}
\label{oscillate_phi}
\phi(t,r) = \phi_o\left(r \right) \,e^{\left(-i \omega t\right)}.
\end{equation}
With this assumption, our goal is then to 
to find $\phi_o\left(r\right)$, $\omega$ and the
metric coefficients such that the spacetime generated by this
matter configuration is static. We begin by considering the problem in
polar-areal coordinates as is done in, for instance, \cite{kaup,hawley,lai}.
The line element in these coordinates takes the form,
\begin{equation}
ds^2 = - \alpha\left(r\right)^2 dt^2 + a\left(r\right)^2 + r^2 d\Omega^2.
\end{equation}
The equilibrium equations in this coordinate system are then given by,
\begin{eqnarray}
\nonumber
a^\prime      &=& \frac{a}{2}      \left\{-    \frac{a^2-1}{r} \right.\\
               && \quad \quad \left. + 4 \pi r \left[ \left(\frac{\omega^2}{\alpha^2}+m^2\right) a^2 \phi_o^2 
	       + \Phi^2_o \right] \right\},\\
\nonumber
\alpha^\prime &=&  \frac{\alpha}{2}\left\{\ \ \frac{a^2-1}{r} \right.\\
               && \quad \quad \left. + 4 \pi r \left[ \left(\frac{\omega^2}{\alpha^2}-m^2\right) a^2 \phi_o^2 
	       + \Phi^2_o \right] \right\}, \label{slicing}\\
\phi^\prime_o &=& \Phi_o,\\
\nonumber
\Phi^\prime_o &=& - \left(1 + a^2 - 4 \pi r^2 a^2 m^2 \phi_o^2  \right) \frac{\Phi_o}{r} \\
               && - \left( \frac{\omega^2}{\alpha^2} - m^2 \right) \phi_o\, a^2.
\end{eqnarray}
where a prime denotes differentiation with respect to $r$.
In order to obtain a solution of this system adapted to the desired physical
situation we provide the following boundary conditions:
\begin{eqnarray}
a\left(0 \right)      &=& 1,\\
\phi_o\left(0 \right) &=& \phi_{c},\\
\Phi_o\left(0 \right) &=& 0,\\
\partial_r \alpha\left(0 \right) &=& 0,\\
\lim_{r\rightarrow \infty}a\left(r \right)      &=& 1,\\
\lim_{r\rightarrow \infty}\phi_o\left(r \right) &=& 0,\\
\lim_{r\rightarrow \infty}\alpha\left(r \right) &=& 1;
\end{eqnarray}
which guarantee regularity at the origin and asymptotic flatness.
For a given value of central value of the field $\phi_c$ these
equations and boundary conditions only admit solutions for a
discrete set of values of $\omega$. In our particular case, we
we are interested on the fundamental (lowest frequency) solution. The
problem is solved by integrating from $r=0$ outwards using a second
order shooting method implemented using the numerical package {\tt
LSODA}\cite{lsoda}. The equations are integrated setting $\alpha(r=0)=1$ initially,
after all the equations are solved, the lapse is rescaled  in order to obtain a function which
asymptotes to $1$ at infinity\footnote{this can be done because
the linearity of the slicing equation (\ref{slicing})}. The  
same rescaling is then performed to the frequency $\omega$.

Once the solution is computed in this coordinate system a change
of coordinates is performed to maximal isotropic ones:
\begin{equation}
ds^2 = \alpha^2\left(\tilde r \right) dt^2 + \psi^4\left(\tilde r \right) \left( d\tilde r^2 + \tilde r^2 d\Omega^2 \right).
\end{equation}
In these coordinates the extension to three dimensions is direct
since the solution is explicitly spatially conformally flat. We
perform the transformation of coordinates as in \cite{lai}. Since
$\alpha\left(\tilde r\right)$ and $\psi\left(\tilde r \right)$ 
(and the scalar field $\phi_o\left(\tilde r \right)$) are
scalar functions of the spacelike hypersurface, and the time-slicing
is not changed, the extension to three-dimensions involves only
writing these functions as functions of $x$, $y$ and $z$ such that
$\tilde r^2 = x^2 + y^2 + z^2$. This task is performed using a $5$
point Lagrange interpolation.

This way we obtain initial data for $g_{ab}$ and $\phi$, the rest of the fields for  the three 
dimensional code are chosen as follows: $Q_{ab}=0$, $\Pi$ is computed from ansatz~(\ref{oscillate_phi}) 
and $D_{iab}$ and $\Phi_i$ are calculated using constraint equations~(\ref{space1}). That completely
defines the initial data for a boson star.


\section{Energy Considerations}
\label{energy_considerations}
To gain some physical insight into these results,
we consider here how the energy density behaves for
the three different cases studied. To simplify the 
discussion, we consider the energy density in flat spacetime, i.e.
we assume a Minkowski metric.
Under this assumption, the energy density for a complex scalar field $\phi$ can be written as
\begin{equation}
\rho = \frac{1}{2}\left[|\Pi|^2 + |\Phi|^2 + m^2 |\phi|^2\right]
\end{equation}
where $\Pi$ and $\Phi$ are the derivatives of $\phi$ defined as before in Eqs.~(\ref{time1}) and (\ref{space1}).
Now, we treat the scalar field as given by the superposition $\phi=\phi_1+\phi_2$ 
with $\phi_1,\phi_2$ describing the different stars we study. The energy density associated 
with this superposition  can be expressed as
\begin{equation}
\rho = \rho_1 + \rho_2 + \Delta
\end{equation}
where $\rho_i$ the energy density that corresponds to the field $\phi_i$ in isolation and
$\Delta$ is the interaction potential which vanishes when
the stars are well separated. The interaction potential can then be written as
\begin{eqnarray}
\nonumber
\Delta = \frac{1}{2} \left[ \bar \Pi_1  \Pi_2  + \Pi_1  \bar \Pi_2 +
                                \bar \Phi_1 \Phi_2 + \Phi_1 \bar \Phi_2 \right.\\
                          \left.+ m^2 \left(\bar \phi_1 \phi_2 + \phi_1 \bar \phi_2\right)
                          \right] ~~.
\end{eqnarray}
We next assume the scalar field assumes the form
\begin{eqnarray}
\phi_1 & = & \phi^0_1(t,x^i) e^{i\omega t} \\
\phi_2 & = & \phi^0_2(t,x^i) e^{i \left( \epsilon \omega t + \theta \right) } 
~~,~~ \epsilon = \pm 1
\end{eqnarray}
where $\phi_1$ represents the complex field configuration associated with
a dynamic boson star and $\phi_2$ represents the field configuration of
the other star. Both stars are characterized by the same frequency $\omega$
but the second star can be a regular boson ($\epsilon=1$, $\theta=0$), a boson in opposition
of phase ($\epsilon=1$, $\theta=\pi$), or an antiboson ($\epsilon=-1$, $\theta=0$). The
profile of each is described by the real field $\phi^0_A(t,x^i)$ which is a assumed to be a slowly
varying function of time so the most dynamical part of the time dependence is represented
by the harmonic factor $e^{i \omega t}$. 

Under these assumptions, a straightforward computation reveals that the effective
interaction potential can be expressed as
\begin{equation}
\Delta =  \Delta_0 ~ \cos \left[ \left( 1 - \epsilon \right) \omega t - \theta  \right] \, .
\end{equation}
Notice that when the two stars are centered in the same position the function $\Delta_0$ is strictly non-negative,
since it is a sum of quadratic terms, 
\begin{eqnarray}
\Delta_{0} &=&   \left[   \bar \Pi^o_1  \Pi^o_2 + \Pi^o_1  \bar \Pi^o_2  \right.\\
\nonumber
                       &&  \left. + \bar \Phi^o_1 \Phi^o_2 + \Phi^o_1 \bar \Phi^o_2 \right.
                            \left.+  m^2  \left(\bar \phi^o_1 \phi^o_2 + \phi^o_1
		        \bar \phi^o_2\right) \right]~~,
\end{eqnarray}
and the behavior of the interaction term for each case considered is governed
by that of $\cos \left[ \left( 1 - \epsilon \right) \omega t - \theta  \right]$.
Thus, for the problems analyzed in the present work, the interaction potential for each case, 
is,
\begin{eqnarray}
\Delta_{\rm B-B  } &=&  \Delta_0  ~~,\\
\Delta_{\rm B-OPB} &=& -\Delta_0 ~~, \nonumber \\ 
\Delta_{\rm B-AB } &=& \Delta_0  \cos ( 2 \omega t) ~~. \nonumber
\end{eqnarray}
Since the contribution of $\Delta_0$ is positive for the boson-boson case, the resulting
gravitational potential associated with this case would be deeper than if not present.
As a result, a merge of the two stars would seem a natural consequence.
On the other hand the $\Delta_0$ contribution has the opposite sign for 
the boson-opposition phase boson case, translating into a gravitational
potential exhibiting a  bump at the center with two minimums around it.
Such a bump would suggest it energetically preferable for the stars not to merge.
Finally, the boson-antiboson interaction potential is governed by a varying function of time.
If the interaction time-scale is much longer than $\simeq 1/\omega$, the interaction term
essentially integrates away to 0. Whether this is the case depends on the collision under
study and, in particular, the stars' momentum as they come close to each other.


\bibliographystyle{prsty}

\begin{thebibliography}{99}

\bibitem{wheeler} J.~A.~Wheeler,
    {\it Phys.~Rev.}, {\bf 97}, 511 (1955).

\bibitem{kaup}  D.~J.~Kaup. Klein-Gordon Geom.
            {\it Phys. Rev.}, {\bf 172}, 1331 (1968).

\bibitem{ruffini} R.~Ruffini, and S.~Bonazzola,
           {\it Phys.~Rev.}, {\bf 187}, 1767 (1969).    
    
\bibitem{jetzer} P.~Jetzer,
           {\it Phys.~Rep.} {\bf 220}, 163 (1992)

\bibitem{mielke} E.~W.~Mielke, and F.~E.~Schunck,
           {\it Class.~Quantum.~Grav.} {\bf 20}, R301 (2003)

\bibitem{dark_matter} E.~W.~Mielke, B.~Fuchs, F.~E.~Schunck
     Proceedings of the 10th Marcel Grossmann Meeting (2006). astro-ph/0608526.

\bibitem{guzman06} F.~S.~Guzman
 Phys.~Rev.~D {\bf 73}, 021501(R), (2006).

\bibitem{lai} K.~W.~Lai, Ph.D. Thesis, University of British Columbia (2004),
gr-qc/0410040.

\bibitem{hawley} S.~H.~Hawley and M.~W.~Choptuik,
 Phys.~Rev.~{\bf D62} (2000) 104024, gr-qc/0007039.
 
\bibitem{guz04} F.~S.~Guzman, 
 Phys.~Rev.~D {\bf 70}, 044033, (2004).

\bibitem{BBDGS05} J.~Balakrishna, R.~Bondarescu, G.~Daues, F.~S.~Guzman and
  E.~Seidel (2006).
 
\bibitem{balakrishna_phd} J.~Balakrishna, Ph.D. Thesis, Washington University (1999),
 gr-qc/9906110.
 
\bibitem{frans} F.~Pretorius, In preparation.
 
\bibitem{Qball}
R.A.~Battye and P.M.~Sutcliffe
"Q-ball Dynamics", Nuclear Physics {\bf 590}, 329 (2000).
 
\bibitem{das}
A.~Das, J.~Math.~Phys.~{\bf 4}, 45 (1963).

\bibitem{Choquet55} Y.~Choquet-Bruhat, Acta Math. {\bf 88} 141 (1955).
       
\bibitem{Frie85} H.~Friedrich, Commun. Math. Phys. {\bf 100} 525 (1985).

\bibitem{Garfinkle} 
D. Garfinkle. Phys.~Rev.~{\bf D65} 044029 (2002).
      
\bibitem{LSKOR06} L.~Lindblom et al.,
        Class.~Quantum.~Grav. {\bf 23}, 447 (2006).

\bibitem{GCHM05} C.~Gundlach, G.~Calabrese, I.~Hinder and
    J.~M.~Martin-Garcia, Class.~Quantum.~Grav. {\bf 22}, 3767 (2005).
	
\bibitem{colpi} M.~Colpi, S.~L.~Shapiro, and I.~Wasserman
 Phys.~Rev.~Lett.{\bf 57} 2485 (1986)

\bibitem{olsson}
 P. Olsson, Math. Comp. {\bf 64}, 1035 (1995); {\bf 64} S23
  (1995); {\bf 64}, 1473 (1995).

\bibitem{gustaffsonkreissoliger} B. Gustafsson, H. Kreiss, and J. Oliger
``Time dependent problems and difference methods'',
John Wiley and Sons, 1995.
	
\bibitem{tadmor}
D. Levy and E. Tadmor,
SIAM Journal on Num. Anal {\bf 40}, 40, (1998).

\bibitem{strand}
B. Strand,
``Summation by parts for finite difference approximations for d/dx'',
Journal of Computational Physics {\bf 110}, 47 (1994).
		
\bibitem{SBP0}
G.~Calabrese, L.~Lehner, D.~Neilsen, J.~Pullin, O.~Reula, O.~Sarbach and M.~Tiglio,
Class.\ Quant.\ Grav.\  {\bf 20}, L245 (2003).

\bibitem{SBP1}
  G.~Calabrese, L.~Lehner, O.~Reula, O.~Sarbach and M.~Tiglio,
  Class.\ Quant.\ Grav.\  {\bf 21}, 5735 (2004).
  
\bibitem{SBP2}
  L.~Lehner, D.~Neilsen, O.~Reula and M.~Tiglio,
  Class.\ Quant.\ Grav.\  {\bf 21}, 5819 (2004).
		
\bibitem{lehnermoreschi}
L. Lehner and O. Moreschi. In preparation (2006).

\bibitem{baumgarteetal}
T. Baumgarte, P. Brady, J. Creighton, L. Lehner, F. Pretorius and R. DeVoe,
``Learning about compact binary merger: the interplay between
numerical relativity and gravitational-wave astronomy''. To be submitted.
	
\bibitem{lsueffects}
M. Anderson, L. Lehner, I. Olabarrieta, C. Palenzuela, S. Liebling, in preparation.

	
\bibitem{berger2}
M.J.~Berger, and J.~Oliger.
``Adaptive Mesh Refinement for hyperbolic partial differential equations,"
J. Comp. Phys. {\bf 53}, 484 (1984).
	   
\bibitem{had1} S.~L.~Liebling, Phys.~Rev.~D {\bf71}, 044019 (2005). gr-qc/0502056
\bibitem{had2} S.~L.~Liebling, Class. Quan. Grav. {\bf 21}, 3995 (2004).gr-qc/0403076

\bibitem{amr} L.~Lehner, S.~L.~Liebling, and O.~Reula,
   Class.~Quantum Grav. {\bf 23}, S421-S446 (2006).

\bibitem{pretoriusphd}
F. Pretorius, ``Numerical simulations of gravitational collapse'',
The University of British Columbia, (2002).

\bibitem{matt_paper}
  M.~Anderson, E.~Hirschmann, S.~L.~Liebling and D.~Neilsen,
  Class.\ Quant.\ Grav.\  {\bf 23}, 6503 (2006)
  [arXiv:gr-qc/0605102].

\bibitem{laipaper} 
D. Choi, K.~W.~Lai, M. Choptuik, E. Hirschmann, S. Liebling and F. Pretorius,
in preparation (2006).
    
\bibitem{damour}
T. Damour. ``The problem of motion in Newtonian and Einsteinian
gravity''. in S.W. Hawking and W. Israel, Eds.``300 Years of
Gravitation''. 128-198. Cambridge University Press. Cambridge, 1987.
  
\bibitem{will}
C. M. Will,
``The Confrontation between General Relativity and Experiment,''
Living Rev. Relativity {\bf 9} 3. URL (cited on 12/08/06):
http://www.livingreviews.org/lrr-2006-3
  
\bibitem{buonano}
  A.~Buonanno, G.~B.~Cook and F.~Pretorius,
  arXiv:gr-qc/0610122.
  
\bibitem{Campanelli:2006gf}
  M.~Campanelli, C.~O.~Lousto and Y.~Zlochower,
  Phys.\ Rev.\ D {\bf 73}, 061501 (2006)
  [arXiv:gr-qc/0601091].
  
    
\bibitem{Baker:2006yw}
  J.~G.~Baker, J.~Centrella, D.~I.~Choi, M.~Koppitz and J.~van Meter,
  Phys.\ Rev.\ D {\bf 73}, 104002 (2006)
  [arXiv:gr-qc/0602026].
  

\bibitem{Bruegmann:2006at}
  B.~Bruegmann, J.~A.~Gonzalez, M.~Hannam, S.~Husa, U.~Sperhake and W.~Tichy,
  arXiv:gr-qc/0610128.  
  
\bibitem{lsoda} L.~R.~Petzold and A.~C.~Hindmarsh, ``LSODA". Computing and Mathematics Research Division,
I-316 Lawrence Livermore National Laboratory, Livermore, CA 94550.

\bibitem{brandt} A.~Brandt
Springer Lecture Notes in Mathematics {\bf 960}, 1 (1982).
      
  

\end{thebibliography}

\end{document}